\documentstyle[psfig,natbib209]{mn-nat}
\newcommand\aj{{AJ}}% 
\newcommand\araa{{ARA\&A}}% 
\newcommand\apj{{ApJ}}% 
\newcommand\apjl{{ApJ}}% 
\newcommand\apjs{{ApJS}}% 
\newcommand\apss{{Ap\&SS}}% 
\newcommand\aap{{A\&A}}% 
\newcommand\aapr{{A\&A~Rev.}}% 
\newcommand\aaps{{A\&AS}}% 
\newcommand\mnras{{MNRAS}}% 
\newcommand\pasp{{PASP}}% 
\newcommand\procspie{{Proc.~SPIE}}% 
\def\ah{\hbox{$^{\rm h}$}}
\def\am{\hbox{$^{\rm m}$}}
\def\Mpc{{\rm\thinspace Mpc}}
\def\kpc{{\rm\thinspace kpc}}
\def\pc{{\rm\thinspace pc}}
\def\s{{\rm\thinspace s}}
\def\y{{\rm\thinspace yr}}
\def\asec{{\rm\thinspace arc second}}
\def\kms{\hbox{$\km\s^{-1}\,$}}
\def\kmsmpc{\hbox{$\km\s^{-1}\,\Mpc^{-1}$}}
\def\km{{\rm\thinspace km}}
\def\lo{\hbox{${\rm\thinspace L}_{\odot}$}}
\def\mo{\hbox{${\rm\thinspace M}_{\odot}$}}
\def\moy{\hbox{${\rm\thinspace M}_{\odot}\y^{-1}$}}
\def\Mb{\hbox{{\rm M}$_B$}}
\def\Bj{\hbox{$b_J$}} 
\def\bj{\hbox{$b_J$}}
\def\Bt{\hbox{$B_T$}} 
  
\def\ha{\hbox{{\rm H}$\alpha$}}
\def\Ha{\hbox{{\rm H}$\alpha$}}
\def\nii{\hbox{\rm [N\thinspace II]}}
\def\m32{\hbox{{\rm M\thinspace 32}}}
\def\iraf{{\small IRAF}}
\def\dofibers{{\small DOFIBERS}}
\def\rvsao{{\small RVSAO}}
\def\one_wide{8.5cm}
\def\two_wide{17.8cm}
\def\ngal{675}
\def\nred{526}
\def\nreds{10}  %stars
\def\nredm{108} %member galaxies
\def\nredb{408} %background galaxies
\def\nredg{516} %no of galaxies = nredm + nredb

\def\cdntot{131}
\def\cdnobs{97}
\def\cdnred{76}

\begin{document}

\title[Dwarf galaxies in Fornax]{The evolution and star formation of dwarf
galaxies in the Fornax Cluster}

\author[Drinkwater et al.]  {M.J.\ Drinkwater$^{1,5}$, M.D.\ Gregg$^{2,3}$,
B.A.\ Holman$^1$, M.J.I.\  Brown$^{1,4}$\\  
$^1$School of Physics, University of Melbourne, Victoria
3010, Australia\\
$^2$Department of Physics, University of California at Davis, Davis,
CA 95616, USA\\
$^3$Institute for Geophysics and Planetary Physics, Lawrence Livermore
National Laboratory, L-413, Livermore, CA 94550, USA\\
$^4$National Optical Astronomy Observatory,
950 North Cherry Avenue,
P.O. Box 26732, Tucson, Arizona 85726, USA\\
$^5$mjdrin@unimelb.edu.au
}

\date{Version 2; revised.} \maketitle

\begin{abstract}
We present the results of a spectroscopic survey of \ngal\ bright
($16.5<\Bj<18$) galaxies in a 6 degree field centred on the Fornax
cluster with the FLAIR-II spectrograph on the UK Schmidt Telescope.
Three galaxy samples were observed: compact galaxies to search for new
blue compact dwarfs, candidate \m32-like compact dwarf ellipticals,
and a subset of the brightest known cluster members in order to study
the cluster dynamics.  We measured redshifts for \nredg\ galaxies of
which \nredm\ were members of the Fornax Cluster. Defining dwarf
galaxies to be those with $\bj\geq15$ ($\Mb\geq-16.5$), there are a
total of 62 dwarf cluster galaxies in our sample. Nine of these are
new cluster members previously misidentified as background
galaxies. The cluster dynamics show that the dwarf galaxies are still
falling into the cluster whereas the giants are virialised.

We classified the observed galaxies as late-type if we detected \ha\
emission at an equivalent width greater than 1 \AA. The spectra were
obtained through fixed apertures, so they reflect activity in the
galaxy cores, but this does not significantly bias the classifications
of the compact dwarfs in our sample. The new classifications reveal a
higher rate of star formation among the dwarf galaxies than suggested
by morphological classification: 35 per cent have significant \ha\
emission indicative of star formation but only 19 per cent were
morphologically classified as late-types.

The star-forming dwarf galaxies span the full range of physical sizes
and we find no evidence in our data for a distinct class of
star-forming blue compact dwarf (BCD) galaxy. The distribution of
scale sizes is consistent with evolutionary processes which transform
late-type dwarfs to early-type dwarfs.  The fraction of dwarfs with
active star formation drops rapidly towards the cluster centre: this
is the usual density-morphology relation confirmed here for dwarf
galaxies. The star-forming dwarfs are concentrated in the outer
regions of the cluster, the most extreme in an infalling
subcluster. We estimate gas depletion time scales for 5 dwarfs with
detected H~I emission: these are long (of order $10^{10}\y$),
indicating that an active gas removal process must be involved if they
are transformed into gas-poor dwarfs as they fall further into the
cluster. Finally, in agreement with our previous results, we find no
compact dwarf elliptical (\m32-like) galaxies in the Fornax Cluster.

\end{abstract}
%\keywords{}

%%%%%%%%%%%%%%%%%%%%%%%%%%%%%%%%%%%%%%%%%%%%%%%%%%%%%%%%%%%%%%%%%%%%%%%%%%%%
\section{Introduction}

Our understanding of the role of dwarf galaxies in clusters has long
been dominated by photographic studies of the nearest clusters.  The
major surveys of the Virgo cluster in the North
\citep*{bin1984,bin1985,bin1993} and the Fornax cluster in the South
\citep*{fer1988,fer1989} have established the presence of large
populations of dwarf galaxies in these clusters.  Lacking redshift
information, these works have had to rely on image morphology for
statistical separation of cluster members from the large contamination
of background objects.

When the Fornax Cluster Catalog \citep{fer1989} was compiled, only 85
of the 340 likely cluster members had measured redshifts; of the whole
FCC catalogue of 2678 galaxies, only 112 (4 per cent) had redshifts.
Since then, a number of small-scale spectroscopic surveys have been
made of the cluster \citep*{hel1994,bur1996,hil1999b}, but even as of
late 1999, NED\footnote{The NASA/IPAC Extragalactic Database (NED) is
operated by the Jet Propulsion Laboratory, California Institute of
Technology, under contract with the National Aeronautics and Space
Administration.}  still lists only about 120 objects within a
$5^{\circ}$ radius of NGC1399 with velocities that would place them in
the cluster. In this paper we describe a large new spectroscopic
investigation of the Fornax cluster designed to add radial velocity
membership, spectral classification, and dynamical information to the
existing knowledge of the cluster.  Our study is based on data from
the FLAIR-II spectrograph on the UK Schmidt Telescope.  We measured
the redshifts of over 500 galaxies brighter than $\bj=18$ in the
direction of Fornax, of which 108 are members of the cluster. This
study is complementary to the {\em Fornax Cluster Spectroscopic
Survey} (FCSS) \citet{dri2000} which is using the Two degree Field
spectrograph on the Anglo-Australian Telescope to make a deeper survey
of a smaller region towards the centre of the cluster. Unlike the
current study, the FCSS will measure {\em all} objects in its target
magnitude range ($16.5<\bj<19.7$), irrespective of morphology
(i.e. both `stars' and `galaxies') in order to sample the largest
possible range of surface brightness.

A number of issues regarding the Fornax dwarf population can be
addressed definitively only through spectroscopic information to
determine cluster membership.  One interesting question is the
possible existence of a population of very compact dwarfs that may
have been misclassified as background galaxies in the FCC.
\citet{dri1996} searched unsuccessfully for additional blue compact
dwarf (BCD) galaxies in the Virgo Cluster; any compact Virgo cluster
galaxies must be fainter than their limit of \bj = 17.6.  The Fornax
cluster has a much lower fraction of late-type dwarfs than Virgo
\citep{fer1988}, and its core galaxy density is about twice that of
Virgo \citep{fer1989b}, offering the chance for an interesting
comparison of two dwarf galaxy populations in rather different
environments.  The initial results of our new spectroscopic survey
have already revealed a number of new compact dwarf galaxy members,
both early and late-type \citep[][Paper~I hereafter]{dri1998}. In this
paper we present the rest of our observations with the analysis of a
well-defined sample of compact galaxies.

The FCC lists a total of 131 objects as candidate ``\m32-like'' compact
dwarf ellipticals (cdE).  \citet{fab1973} suggested that high
surface-brightness low-luminosity elliptical galaxies (like \m32\ ) are
formed by tidal stripping of brighter ellipticals in rich
environments, but more recent models find that the amount of mass lost
by this process in a Hubble time is too small to produce compact dwarf
ellipticals from larger galaxies \citep{nie1987}.  To test the tidal
stripping hypothesis, we observed 97 of the possible cdEs in Fornax to
determine if they were cluster members.  Whilst there was considerable
overlap between these and our compact sample described above, the
\m32-like sample also included a number of galaxies with larger scale
sizes not included in the compact sample. It should be possible to
distinguish galaxies like \m32\  from normal dwarf galaxies on the basis
of their profiles. We would expect \m32-like bulges to have de
Vaucouleurs profiles \citep{mic1991}, whereas most dwarf galaxies have
exponential profiles \citep{fer1994}. Our photographic imaging data is
not able to resolve differences at this level however, hence the
overlap in the two samples.  In Paper~1 we reported that only one of
the \m32\  candidates we observed was a cluster member, but that it had
an emission line spectrum inconsistent with a cdE identification. We
concluded that tidal stripping had not produced large numbers of cdEs
in Fornax \citep{hol1995,dri1997b}. In this paper we expand our
discussion of the \m32\  candidate galaxies in the context of complete
samples.

Our measurements of candidate compact cluster galaxies detected a
majority of background galaxies, interesting in their own right for
large scale structure information. In addition to the new compact
cluster members discovered (Paper~I), we also
observed some 100 known cluster galaxies in order to investigate star
formation rates and the dynamics of the cluster, particularly
subclustering \citep[][Paper~III]{dri2001b}. In this paper we adopt a
cluster distance of 20\Mpc\  \citep[][but see discussion in
Paper~III]{mou2000} corresponding to a distance modulus of 31.5 mag.

In Section~\ref{sec_sample} we describe how the galaxies were selected
and in Section~\ref{sec_obs} we describe the observations and redshift
measurements. We discuss the structure of the cluster in
Section~\ref{sec_structure} and in Section~\ref{sec_starf} we discuss
the star formation process in the dwarf cluster galaxies. The compact
dwarf elliptical population is discussed in Section~\ref{sec_compacts}
and we summarise all our results in Section~\ref{sec_summary}.  We
list the background galaxies we observed in the Appendix.

%%%%%%%%%%%%%%%%%%%%%%%%%%%%%%%%%%%%%%%%%%%%%%%%%%%%%%%%%%%%%%%%%%%%%%%%%%%%
\section{Sample Selection}
\label{sec_sample}

The selection criteria for the objects we observed were set by our
three overlapping science goals, subject to the capacity of the
FLAIR-II spectrograph on the UK Schmidt Telescope of the
Anglo-Australian Observatory \citep{par1995}.  This is a
multi-object fibre spectrograph with a square field approximately 6
degrees across.  The limiting magnitude of the system is about
\bj=17.5 and each observation can take 70 or 90 spectra depending on
the plate holder used.  It is ideally suited to the measurement of the
brighter galaxies in the Fornax cluster which has a core radius of
$0.7^{\circ}$ (FCC).

\begin{figure}
\psfig{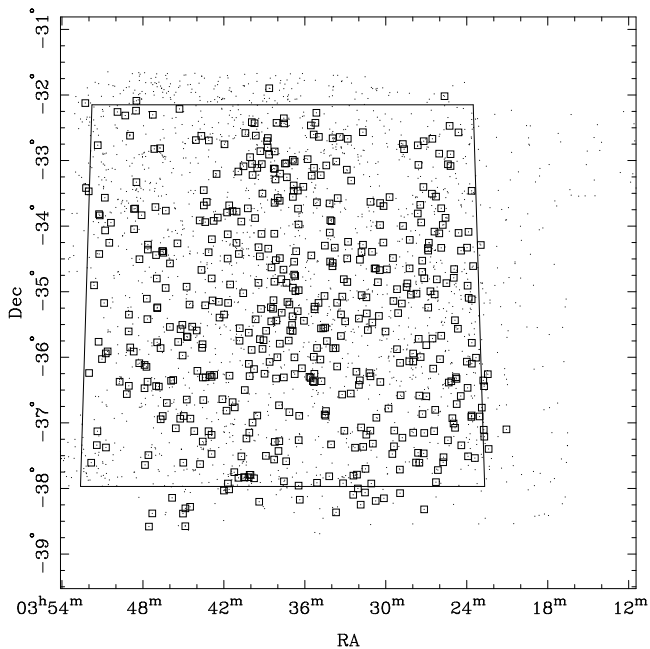}
\centering
\caption{Distribution on the sky of all FCC galaxies (small points),
and galaxies observed with FLAIR-II (squares). The region defined by the
APM sample is also shown.}
\label{fig_sky}
\end{figure}

Most of out observations were made centred on the standard UK Schmidt
sky survey field termed ``F358'' (03\ah37\am55\fs9,
$-$34\degr50\arcmin14\arcsec\ J2000).  As seen in Fig.~\ref{fig_sky},
the cluster is offset from field F358, so some observations were
centred on a second field displaced about half a degree southwest
(03\ah36\am55\fs1 $-$35\degr30\arcmin11\arcsec\ J2000). The
overlapping samples we observed are described in the rest of this
section.

\subsection{Compact galaxies}

The primary goal of the project was to search for new cluster compact
galaxies, both red and blue, so we concentrated efforts on galaxies
with a compact morphology that were {\em not} classified as cluster
members in the FCC. These were selected on the basis of their image
parameters measured from the two Schmidt plates described
above. Although the targets were initially selected on the basis of
compact image morphologies measured from the first Schmidt plate (as
described below), many additional targets were later added as the
second plate was used and additional ``\m32-like'' candidates were
included.  We can therefore define three different samples of compact
galaxies: a statistical sample with defined limits and completeness
from a single plate; the list of all \m32\ candidates observed; and a
heterogeneous list of both of these plus any other compact galaxies
observed.

The first sample was defined by image parameters measured by the APM
(Automated Plate Measuring facility, Cambridge) from the UK Schmidt
\bj\ survey plate of field F358. See \citet{irw1994} for more details
of the APM image catalogue.  The APM image catalogue lists image
positions, magnitudes and morphological classifications (as `star',
`galaxy', `noise', or `merged') measured from both the blue (\Bj) and
red survey plates.  The `merged' image classification indicates two
overlapping images: at the magnitudes of interest for this project,
the merged objects nearly always consisted of a star overlapping a
much fainter galaxy.  The APM magnitudes are calibrated for unresolved
(stellar) objects, so for most purposes we use magnitudes estimated by
fitting exponential disk profiles to the APM image parameters
calculated by \citet{dav1988} and \citet{irw1990}
\citep[see][]{mor1999a}. We refer to these as Davies magnitudes. This
process also measures the exponential scale lengths of the images. For
galaxies brighter than \bj=13 we have used magnitudes from
\citet{fer1989}. There is no evidence for any significant offset
between the Davies and FCC magnitudes \citep{dri2000}.  In one
specific case (FCC~39) there seems to be a problem with both the FCC
and Davies magnitudes being too faint, so we have adopted the value of
\bj=13.6 given by \citet{mad1990}.

\begin{figure}
\psfig{file=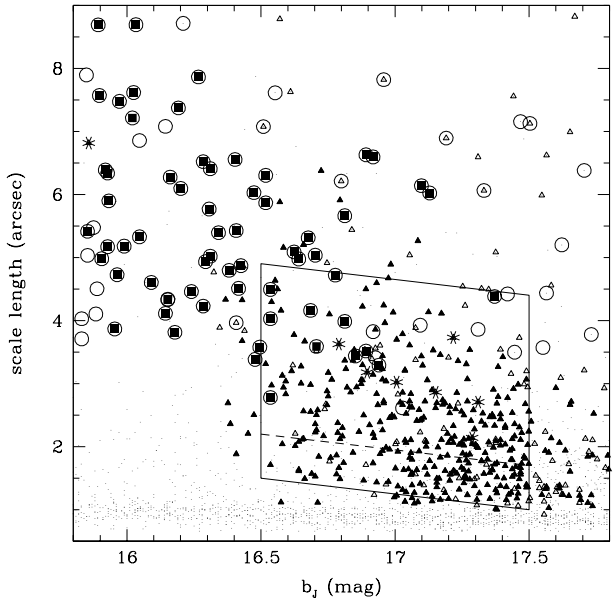,angle=0,width=\one_wide}
\centering
\caption{Distribution of \bj\  magnitudes and scale length of all
resolved objects in the APM catalogue (small points), galaxies
observed with FLAIR-II (open triangles; those with redshifts are
filled triangles and cluster members are indicated by large filled
squares) and galaxies listed as members in the FCC (large
circles). New members of the cluster discovered with FLAIR are
indicated by asterisks. The parallelogram defines the compact galaxy
sample discussed in the text. Objects in the parallelogram above the
dashed line form the more conservative sample.  Note that for the
definitions of these samples, the magnitudes were taken from the APM
catalogue, rescaled to approximate the calibrated Davies magnitudes
used elsewhere.}
\label{fig_morph}
\end{figure}

\begin{figure}
\psfig{file=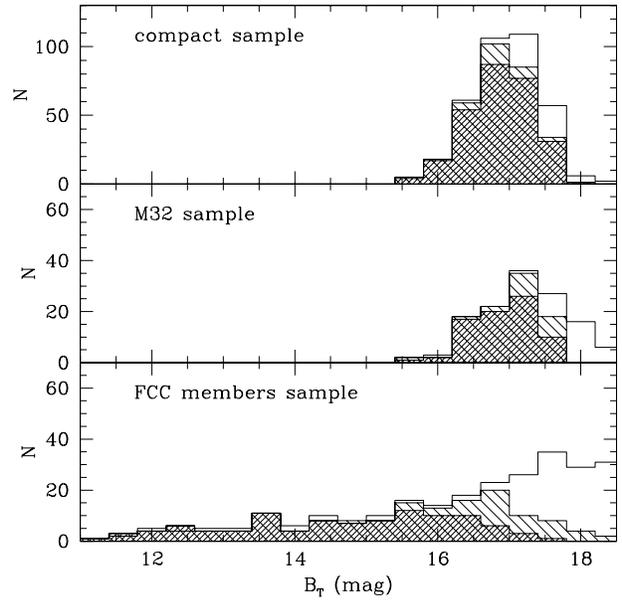,angle=0,width=\one_wide}
\centering
\caption{Distributions of \Bt magnitudes (from the FCC) of all galaxies
from each of the three samples observed: compact galaxies, \m32\ 
candidates from the FCC and definite cluster members from the FCC.  In
each case the unshaded histogram shows the parent sample, the shaded
histogram shows the galaxies observed, and the solid histogram those
galaxies with measured redshifts.  Note the different vertical scale in
the top panel.}
\label{fig_sample}
\end{figure}

We defined the compact galaxy sample by scale lengths and magnitudes
as shown in Fig.~\ref{fig_morph}. To optimise our measurement for the
compact objects we used the APM (stellar) magnitudes for this
selection, linearly scaled to agree with the Davies magnitudes over
this range. The compact galaxy sample is defined in the magnitude
range of $16.5<\Bj<17.5$ by the parallelogram shown in the
Figure. Basically it includes all galaxies and possible galaxies
(``merged'' objects) in that magnitude range with sizes smaller than
the galaxies classified as cluster members in the FCC. This compact
subsample consists of 815 objects of which 25 are FCC members, 400 are
FCC background galaxies and the remainder (390) are not listed in the
FCC, presumably because they are smaller than the FCC limiting
diameter of 17 arc seconds. We successfully observed 306 or 38 per
cent of the objects in this sample. Most of the objects not observed
were classified as ``merged'' and were not measured because they were
clearly dominated by a star with visible diffraction spikes.  The
magnitude distribution of the compact sample is compared with the
other samples observed in Fig.~\ref{fig_sample}.

None of the new cluster members were found amongst the smallest
objects in the compact sample: these were mostly merged objects
dominated by a star. We therefore defined a second, more conservative
compact sample by excluding the smallest objects below the dashed line
in Fig.~\ref{fig_morph}. This second sample has a higher completeness,
comprising 352 compact objects (25 FCC members, 272 FCC background
galaxies and 55 not listed in the FCC) of which we measured 211 (60
per cent) to find 7 new members.

Our second compact galaxy sample consisted of the \cdntot\   galaxies
listed in the FCC as possible cdE ``\m32-like'' candidates.  We
observed \cdnobs\   of these, obtaining \cdnred\   reliable redshifts.
Fig.~\ref{fig_sample} shows that these objects were observed to a
magnitude limit of $\Bt=17.8$.

Our total compact galaxy sample consisted of the above two samples
plus additional compact objects with similar properties. From this
sample we measured a total of 437 redshifts.

\subsection{Cluster members}

We also attempted to observe as many as possible of the cluster
members from the FCC in order to have a good sample for studies of the
cluster dynamics (Paper~III).  These were measured 
 with a higher spectral dispersion to give more accurate
velocities. The Fornax cluster extends well beyond the limits of a
single Schmidt plate; the FCC includes galaxies in a slightly larger
region than defined by our APM sample as is shown in
Fig.~\ref{fig_sky}. This Figure shows that we observed some galaxies
beyond the APM sample.  Fig.~\ref{fig_sample} shows that the member
galaxies observed were as faint as $\Bt=18$, but being of generally
low surface brightness the success rate fainter than $\Bt=16.6$ was
lower than for the other samples.

%%%%%%%%%%%%%%%%%%%%%%%%%%%%%%%%%%%%%%%%%%%%%%%%%%%%%%%%%%%%%%%%%%%%%%%
%%%%%%%%%%%%%%%%%%%%%%%%%%%%%%%%%%%%%%%%%%%%%%%%%%%%%%%%%%%%%%%%%%%%%%%
\section{Observations}
\label{sec_obs}

\subsection{FLAIR-II Observations}
\label{sec_flair}

We observed Fornax for a total of 6 observing runs. The first 5 used
the low-dispersion 250B grating (coverage 3670-7230\AA, resolution
13\AA), chosen to identify as many galaxies as possible at all
redshifts.  These observations were severely affected by bad weather,
but by the end of the fifth run we had measured most of the compact
galaxies within our magnitude limits.  For the sixth run we observed
FCC-classified members (most of which we had not yet observed) with
the medium-dispersion 600V grating (coverage 5150--6680\AA, resolution
5.3\AA) to measure more accurate velocities and study the dynamics of
the cluster.  A log of our observing runs is shown in
Table~\ref{tab_dates}.

\begin{table}
\caption{Details of our FLAIR-II observations.\label{tab_dates}}
\begin{tabular}{llrl}
\hline
  Run&  Date&$v_{helio}$ & grating \\  %   name&
    &              & (\kms)        \\  %       &
                                  \hline        
   1&  1992 October 25  &   1.4  & 250B \\  %  Fnx-a&
   1&  1992 October 26  &   1.1  & 250B \\  %  Fnx-b&
   2&  1993 November 14 & $-$5.3 & 250B \\  %  Fnx-c&
   2&  1993 November 15 & $-$5.6 & 250B \\  %  Fnx-d&
   2&  1993 November 16 & $-$5.9 & 250B \\  %  Fnx-e&
   3&  1994 November 06 & $-$2.7 & 250B \\  %  Fnx-f&
   4&  1995 November 22 & $-$7.7 & 250B \\  %  Fnx-g&
   4&  1995 November 26 & $-$8.9 & 250B \\  %  Fnx-h&
   5&  1996 December 08 &$-$12.5 & 250B \\  %  Fnx-i&
  5&  1996 December 09 &$-$12.7 & 250B \\  %  Fnx-j&
  6&  1997 October 31  & $-$0.7 & 600V \\  %  Fnx-k&
  6&  1997 November 01 & $-$1.0 & 600V \\  %  Fnx-l& 
  6&  1997 November 08 & $-$3.3 & 600V \\  %  Fnx-m&
\hline
\end{tabular}

Notes: $v_{helio}$ is the correction applied to convert our observed
velocities to heliocentric values.
\end{table}

\subsection{Data Reduction and Velocity Measurement} 

The data were reduced using the \iraf\  \citep{tod1993} \dofibers\
package.  We used the \iraf\  add-on package \rvsao\  \citep{kur1998} to
measure the galaxy redshifts by cross-correlation with template
spectra. All velocities were then corrected to heliocentric values
using the offsets listed in Table~\ref{tab_dates}.

We checked our velocity errors using measurements of 52 galaxies that
were observed twice.  The mean and rms\  values of the velocity
differences are shown in Table~\ref{tab_errors} for the whole sample
and then for two sets of subsamples split according to grating and
also by magnitude.  For comparison, we also give the \rvsao\  estimated
mean velocity errors for each sample: these are based on a formal
statistical calculation.  The results shown in the table are very good
considering the relatively low dispersion and signal-to-noise of our
spectra. The velocity resolutions of the 250B and 600V gratings are
650\kms\  and 265\kms\  respectively.  In particular, the error
estimates from \rvsao\  are shown to be realistic, underestimating the
empirically-determined errors by only around 10 per cent.  The
improvement with the 600V grating is less than expected, but this is
based on a very small number of repeated observations of poor spectra
at this dispersion.  Note that the rms\  values discussed here are
for the difference of two measurements and are a factor of $\sqrt 2$
larger than the rms\  for a single measurement. We have therefore
adopted the velocity uncertainties measured by \rvsao.

As a further test of our velocity measurements we compared them to
previous measurements in the literature. We found redshift
measurements of 102 galaxies in common with our sample in NED (as of
1999 November 23).  The mean velocity difference (FLAIR$-$NED) was
$\overline{\Delta v} = -22\pm11\kms$ and $\sigma_{\Delta v}=108\kms$,
demonstrating that there were no large systematic errors in our
measurements.

\begin{table}
\caption{Velocity differences for repeated observations\label{tab_errors}}
\begin{tabular}{lllll}
\hline
sample&$N$&$\overline{\Delta v}$&$\sigma_{\Delta v}$&$\overline{\sigma_v}$ \\
    &              & (\kms)    & \multicolumn{2}{r}{(\kms)}   \\  % 
                                  \hline
all  & 52 & 10 & 67 & 59 \\
600V &  9 & 32 & 45 & 44 \\
250B & 43 &  4 & 69 & 60 \\
$\Bt<16$&22&26 & 56 & 53 \\
$\Bt>16$&30& 2 & 74 & 62 \\
\hline
\end{tabular}

Notes: $\overline{\Delta v}$ and $\sigma_{\Delta v}$ are the mean and
rms velocity differences from repeated observations.  For each sample
$\overline{\sigma_v}$ is the mean statistical uncertainty based on the
\rvsao\   errors. These rms\   values are for the difference of two
measurements and are $\sqrt 2$ larger than the rms\   for a single
measurement.
\end{table}

In the course of our 6 observing runs, we observed a total of \ngal\   
objects and obtained reliable redshifts for \nred\  of them.  We found
that \nreds\  of these objects had absorption line spectra and
velocities consistent with Galactic stars.  These were merged objects
consisting of a star superimposed on a background galaxy. These were
included in the compact sample in case the ``star'' was the core of
the galaxy, but in each case it was shown to be in the foreground and
dominated the spectrum so the background galaxy could not be
measured. These objects are listed for reference in
Table~\ref{tab_star} and are not discussed further in this paper.

Our final sample consists of \nredm\  confirmed cluster galaxies with
successfully measured redshifts and is listed in Table~\ref{tab_cat}.
The \nredb\  confirmed background galaxies are listed in
Table~\ref{tab_back}. Note that our position for the galaxy FCC~13 is
taken from \cite{lov1996b} as the FCC position is in error.

\begin{table}
\caption{Objects with spectra dominated by light from a foreground
Galactic star\label{tab_star}}	
\begin{tabular}{crrrr}
\hline
 RA (J2000) Dec            & \bj &$cz$&$\Delta cz$&FCC \\
                           & (mag)   &\multicolumn{2}{r}{(\kms)}& \\  
                                  \hline
 03:30:10.12  $-$33:41:35.8& 17.40 &  193&  64&    81 \\
 03:31:26.86  $-$37:10:48.5& 16.60 &   41&  23&  B723 \\
 03:33:10.30  $-$37:55:11.9& 17.30 &$-$24&  27&  B849 \\
 03:37:22.01  $-$34:51:54.1& 16.90 &   86&  43&    -- \\   
 03:38:15.18  $-$34:50:55.2& 15.50 &   39&  12& B1239 \\
 03:39:19.56  $-$35:43:28.9& 16.80 & $-$5&  60&   223 \\
 03:44:51.01  $-$38:18:14.0& 16.70 &$-$42&  34& B1781 \\
 03:47:39.18  $-$36:22:19.3& 16.70 &   74&  54&    -- \\
 03:49:18.67  $-$32:18:47.3& 17.10 &   99&  56& B2112 \\
 03:50:48.11  $-$34:03:58.4& 16.10 &    1&  38&    -- \\
\hline 
\end{tabular}           
\end{table}

\begin{table*}
\caption{Catalogue of all Fornax Cluster members with
redshifts measured with FLAIR-II\label{tab_cat}}
{\tiny
\begin{tabular}{rcrrrrrrrr}
FCC&RA (J2000) Dec & \multicolumn{2}{c}{cz (\kms) $\sigma_{cz}$} 
&EW(\ha)(\AA) & T-type & \bj (mag)
& SB (mag$\asec^{-2}$)&$\alpha$ (\asec)&P1(1)\\
\hline
 ~13     &03:21:03.7 $-$37:05:58 & 1792  &  25 &   11.8&   5&  12.0&   ---&  --- &  0.01 \\ 
 ~19     &03:22:22.9 $-$37:23:50 & 1497  &  47 & $-$0.7&$-$5&  15.2&   ---&  --- &  0.01 \\ 
 ~21     &03:22:42.2 $-$37:12:36 & 1659  &  17 &    1.0&$-$1&  09.4&   ---&  --- &  0.01 \\ 
 ~22     &03:22:44.6 $-$37:06:16 & 1980  &  18 &   15.4&   1&  11.9&   ---&  --- &  0.01 \\ 
 ~26     &03:23:37.3 $-$35:46:43 & 1718  &  46 &    3.5&$-$1&  16.4&  22.3&  6.1 &  0.18 \\ 
 ~28     &03:23:54.3 $-$37:30:33 & 1381  &  28 &    9.3&   9&  13.5&  18.4&  3.9 &  0.02 \\ 
 ~29     &03:23:56.3 $-$36:27:57 & 1244  &  19 &   18.6&   1&  11.8&   ---&  --- &  0.01 \\ 
 ~32     &03:24:52.5 $-$35:26:08 & 1318  &  26 &    9.6&$-$5&  16.1&  21.8&  5.4 &  0.17 \\ 
 ~33     &03:24:58.5 $-$37:00:34 & 1994  &  19 &    4.0&   7&  14.2&  21.4&  6.3 &  0.02 \\ 
 ~35     &03:25:04.1 $-$36:55:40 & 1797  &  15 &    175&   9&  16.1&  21.3&  4.2 &  0.02 \\ 
 ~37     &03:25:09.1 $-$36:21:55 & 1785  &  59 &    3.8&   5&  13.8&  21.8&  7.1 &  0.04 \\ 
 ~36     &03:25:12.2 $-$32:54:10 & 2325  &  90 &    1.3&$-$5&  16.2&  21.2&  4.0 &  1.00 \\ 
 ~39     &03:25:19.8 $-$36:23:05 &  991  &  13 &   23.3&   7&  13.6&  21.6&  6.1 &  0.02 \\ 
 ~43     &03:26:02.5 $-$32:53:42 & 1323  &  17 & $-$1.7&$-$5&  15.1&  21.6&  7.6 &  1.00 \\ 
 ~46     &03:26:25.1 $-$37:07:39 & 2220  &  32 &    2.1&$-$5&  16.5&  21.3&  3.6 &  0.06 \\ 
 ~47     &03:26:32.1 $-$35:42:49 & 1473  &  23 & $-$0.6&$-$4&  13.3&  21.3&  8.7 &  0.18 \\ 
 ~55     &03:27:18.1 $-$34:31:34 & 1279  &  17 & $-$0.8&$-$1&  15.5&  20.6&  4.1 &  0.95 \\ 
 B~470   &03:27:33.8 $-$35:43:04 &  723  &  79 &    0.3&   0&  17.5&  21.1&  2.1 &  0.12 \\ 
 ~62     &03:27:58.3 $-$37:08:57 & 1813  &  13 &   61.6&   4&  12.6&  15.9&  3.7 &  0.46 \\ 
 ~67     &03:28:48.8 $-$35:10:49 & 1345  &  44 &    2.8&   5&  14.2&  21.1&  5.5 &  0.85 \\ 
 ~76     &03:29:43.4 $-$33:33:26 & 1808  &  14 &    6.1&  10&  16.3&  22.3&  6.4 &  1.00 \\ 
 ~83     &03:30:35.2 $-$34:51:19 & 1477  &  11 &    0.4&$-$4&  12.3&   ---&  --- &  1.00 \\ 
 ~88     &03:31:08.1 $-$33:37:47 & 1924  &  15 &    0.1&   3&  11.8&   ---&  --- &  1.00 \\ 
 ~90     &03:31:08.2 $-$36:17:24 & 1813  &  15 &   11.2&$-$4&  15.6&  21.8&  6.9 &  1.00 \\ 
 ~95     &03:31:24.8 $-$35:19:51 & 1275  &  26 & $-$0.5&$-$1&  15.2&  20.4&  4.3 &  1.00 \\ 
 B~729   &03:31:32.5 $-$38:03:43 & 1676  &  31 &    6.8&$-$5&  16.5&   ---&  --- &  1.00 \\ 
 ~100    &03:31:47.7 $-$35:03:05 & 1660  &  31 & $-$0.4&$-$5&  15.9&  22.4&  7.9 &  1.00 \\ 
 ~102    &03:32:10.8 $-$36:13:13 & 1723  &  61 &   13.0&  10&  16.8&  22.2&  5.0 &  1.00 \\ 
 ~106    &03:32:47.8 $-$34:14:19 & 2064  &  35 & $-$1.4&$-$1&  16.3&  22.3&  6.3 &  1.00 \\ 
 ~113    &03:33:06.8 $-$34:48:32 & 1365  &  23 &   10.1&   5&  15.5&  22.2&  8.7 &  1.00 \\ 
 ~119    &03:33:33.9 $-$33:34:23 & 1384  &  40 & $-$0.1&$-$1&  15.7&  21.2&  5.0 &  1.00 \\ 
 ~121    &03:33:36.3 $-$36:08:28 & 1446  &  12 &   43.5&   4&  10.2&   ---&  --- &  1.00 \\ 
 B~904   &03:33:56.2 $-$34:33:43 & 2254  &  56 & $-$0.1&$-$5&  17.4&  21.7&  2.9 &  1.00 \\ 
 B~905   &03:33:57.2 $-$34:36:43 & 1242  &  23 &   18.2&  10&  17.7&  22.5&  3.7 &  1.00 \\ 
 ~136    &03:34:29.5 $-$35:32:46 & 1217  &  24 & $-$1.1&$-$5&  16.3&  21.6&  4.8 &  1.00 \\ 
 ~135    &03:34:30.9 $-$34:17:51 & 1381  &  49 & $-$0.5&$-$5&  16.5&  22.3&  5.8 &  1.00 \\ 
 ~139    &03:34:57.4 $-$32:38:22 & 1752  &  24 &   13.6&   9&  15.3&  20.2&  3.8 &  1.00 \\ 
 ~143    &03:34:59.2 $-$35:10:15 & 1364  &  16 & $-$1.1&$-$4&  14.5&  20.9&  7.6 &  1.00 \\ 
 ~147    &03:35:16.9 $-$35:13:39 & 1299  &  17 &   -0.8&$-$4&  12.0&   ---&  --- &  1.00 \\ 
 ~148    &03:35:16.9 $-$35:16:01 &  794  &  17 & $-$1.3&$-$1&  13.9&  19.4&  5.0 &  1.00 \\ 
 B~1005  &03:35:20.4 $-$32:36:08 & 1256  &  36 & $-$0.2&$-$1&  13.7&  19.9&  6.8 &  1.00 \\ 
 ~150    &03:35:24.1 $-$36:21:49 & 2031  &  26 & $-$1.1&$-$5&  16.2&  21.4&  4.5 &  1.00 \\ 
 ~153    &03:35:31.1 $-$34:26:49 & 1589  &  10 & $-$0.7&$-$1&  14.2&  19.2&  4.1 &  1.00 \\ 
 ~152    &03:35:33.2 $-$32:27:50 & 1389  &  12 &    7.4&   0&  14.9&  20.7&  5.9 &  1.00 \\ 
 ~161    &03:36:04.1 $-$35:26:34 & 1342  &  16 & $-$1.4&$-$4&  12.6&   ---&  --- &  1.00 \\ 
 ~164    &03:36:12.9 $-$36:09:59 & 1430  &  33 & $-$1.9&$-$5&  17.0&  22.1&  4.2 &  1.00 \\ 
 ~167    &03:36:27.6 $-$34:58:36 & 1953  &  13 &    0.1&   0&  11.3&   ---&  --- &  1.00 \\ 
 ~170    &03:36:31.7 $-$35:17:48 & 1763  &  14 & $-$0.7&$-$1&  12.5&  17.5&  4.0 &  1.00 \\ 
 ~176    &03:36:45.1 $-$36:15:22 & 1399  &  30 & $-$1.1&   1&  13.7&  21.4&  9.6 &  1.00 \\ 
 ~174    &03:36:45.4 $-$33:00:49 & 1645  &  78 & $-$1.2&$-$5&  16.8&  21.8&  4.0 &  1.00 \\ 
 ~179    &03:36:46.4 $-$36:00:02 &  972  &  17 &    4.5&   1&  12.4&   ---&  --- &  1.00 \\ 
 ~177    &03:36:47.5 $-$34:44:22 & 1622  &  21 & $-$0.3&$-$1&  13.2&  20.3&  5.4 &  1.00 \\ 
 B~1108  &03:36:49.7 $-$33:27:39 & 1734  & 137 & $-$0.1&$-$5&  17.8&  22.5&  3.6 &  1.00 \\ 
 ~181    &03:36:53.2 $-$34:56:17 & 1113  &  53 & $-$0.2&$-$5&  17.7&  22.9&  4.4 &  1.00 \\ 
 ~182    &03:36:54.4 $-$35:22:27 & 1657  &  19 & $-$1.6&$-$1&  15.0&  21.3&  7.2 &  1.00 \\ 
 ~184    &03:36:57.0 $-$35:30:29 & 1257  &  12 &    0.1&$-$1&  12.3&   ---&  --- &  1.00 \\ 
 ~188    &03:37:04.6 $-$35:35:24 & 1004  &  80 & $-$1.6&$-$5&  16.1&  22.1&  6.3 &  1.00 \\ 
 ~190    &03:37:09.0 $-$35:11:42 & 1740  &  17 & $-$1.2&$-$1&  13.5&  20.1&  6.4 &  1.00 \\ 
 ~193    &03:37:11.8 $-$35:44:44 &  921  &  12 & $-$1.1&$-$1&  13.2&   ---&  --- &  1.00 \\ 
 ~201    &03:37:53.7 $-$37:16:47 & 1922  & 126 &    0.7&$-$5&  17.3&  23.2&  6.0 &  1.00 \\ 
 ~202    &03:38:06.6 $-$35:26:23 &  808  &  22 & $-$0.7&$-$4&  16.1&  21.5&  4.9 &  1.00 \\ 
 ~203    &03:38:09.3 $-$34:31:06 & 1173  &  30 & $-$0.6&$-$5&  16.8&  22.1&  4.5 &  1.00 \\ 
 ~206    &03:38:13.5 $-$37:17:23 & 1402  &  20 &   14.4&$-$5&  16.9&  22.9&  6.1 &  1.00 \\ 
 ~204    &03:38:13.8 $-$33:07:36 & 1369  &  28 & $-$0.5&$-$5&  15.8&  21.4&  5.3 &  1.00 \\ 
 ~207    &03:38:19.3 $-$35:07:43 & 1419  &  43 &    2.2&$-$5&  16.1&  21.5&  4.9 &  1.00 \\ 
 ~211    &03:38:21.5 $-$35:15:35 & 2276  &  23 & $-$0.1&$-$4&  16.8&  21.6&  3.6 &  1.00 \\ 
 ~213    &03:38:29.3 $-$35:27:07 & 1433  &  26 &    0.1&$-$4&  10.6&   ---&  --- &  1.00 \\ 
 ~219    &03:38:52.2 $-$35:35:42 & 1897  &  21 &   -0.2&$-$4&  10.0&   ---&  --- &  1.00 \\ 
 B~1379  &03:39:55.0 $-$33:03:09 &  745  &  21 &   17.3&$-$5&  17.3&  21.7&  3.0 &  1.00 \\ 
 ~235    &03:40:09.3 $-$35:37:28 & 1974  &  20 &   23.1&  10&  13.4&  22.1&  7.3 &  1.00 \\ 
 ~243    &03:40:27.0 $-$36:29:56 & 1404  &  52 & $-$2.8&$-$5&  16.7&  22.5&  5.7 &  1.00 \\ 
 ~245    &03:40:33.8 $-$35:01:21 & 2124  &  68 & $-$1.1&$-$5&  16.6&  22.7&  6.6 &  1.00 \\ 
 ~249    &03:40:42.1 $-$37:30:38 & 1571  &  18 & $-$0.6&$-$4&  14.1&  19.7&  5.2 &  1.00 \\ 
 ~252    &03:40:50.4 $-$35:44:53 & 1415  &  35 & $-$1.9&$-$5&  16.1&  22.0&  6.0 &  1.00 \\ 
 ~253    &03:40:55.2 $-$37:50:16 & 1677  &  57 & $-$2.0&$-$5&  17.0&  22.4&  4.7 &  1.00 \\ 
 ~255    &03:41:03.5 $-$33:46:43 & 1283  &  25 & $-$1.0&$-$1&  13.9&  19.2&  4.5 &  1.00 \\ 
 ~261    &03:41:21.5 $-$33:46:09 & 1492  &  42 &    4.5&$-$5&  16.7&  22.3&  5.3 &  1.00 \\ 
 ~263    &03:41:32.3 $-$34:53:22 & 1691  &  15 &   32.9&   6&  14.9&  20.2&  4.7 &  1.00 \\ 
 ~266    &03:41:41.3 $-$35:10:12 & 1551  &  39 & $-$1.3&$-$5&  16.0&  20.7&  3.4 &  1.00 \\ 
 ~267    &03:41:45.6 $-$33:47:29 &  834  &  10 &   13.4&   9&  16.4&  22.3&  5.9 &  1.00 \\ 
 B~1554  &03:41:59.5 $-$35:20:53 & 1642  &  52 & $-$0.8&$-$4&  17.7&  21.8&  2.7 &  1.00 \\ 
 ~276    &03:42:19.3 $-$35:23:41 & 1382  &  12 & $-$0.8&$-$4&  12.6&   ---&  --- &  1.00 \\ 
 ~277    &03:42:22.8 $-$35:09:15 & 1613  &  25 & $-$0.4&$-$4&  13.9&  20.0&  5.0 &  1.00 \\ 
 ~278    &03:42:27.3 $-$33:52:14 & 2125  &  30 & $-$1.2&$-$5&  17.4&  22.1&  3.5 &  1.00 \\ 
 ~282    &03:42:45.6 $-$33:55:12 & 1267  &  36 &    8.2&  10&  15.0&  20.1&  4.3 &  1.00 \\ 
 ~285    &03:43:01.9 $-$36:16:16 &  891  &   6 &   22.8&   7&  14.2&  22.5&  9.3 &  1.00 \\ 
 ~288    &03:43:22.8 $-$33:56:20 & 1103  &  37 & $-$1.7&$-$5&  16.8&  22.0&  4.5 &  1.00 \\ 
 ~290    &03:43:37.1 $-$35:51:17 & 1392  &  20 & $-$0.4&   5&  12.4&   ---&  --- &  1.00 \\ 
 ~296    &03:44:32.9 $-$35:11:45 &  816  & 122 & $-$0.9&$-$5&  16.7&  22.3&  5.0 &  1.00 \\ 
 ~298    &03:44:44.4 $-$35:41:01 & 1719  &  36 & $-$1.1&$-$5&  17.1&  21.7&  3.4 &  1.00 \\ 
 ~299    &03:44:58.6 $-$36:53:42 & 2151  &  40 &   29.7&   7&  17.9&  22.5&  3.3 &  1.00 \\ 
 ~301    &03:45:03.6 $-$35:58:22 & 1005  &  19 & $-$0.9&$-$4&  14.6&  20.2&  5.2 &  1.00 \\ 
 ~302    &03:45:12.3 $-$35:34:14 &  806  &  23 &   20.2&   8&  17.1&  23.2&  6.5 &  1.00 \\ 
 ~303    &03:45:14.1 $-$36:56:12 & 1980  &  31 & $-$2.9&$-$5&  15.5&  21.9&  7.4 &  1.00 \\ 
 ~305    &03:45:33.8 $-$37:04:58 & 1228  &  25 & $-$1.2&$-$5&  16.8&  22.3&  5.1 &  1.00 \\ 
 ~306    &03:45:45.4 $-$36:20:45 &  898  &  14 &   34.7&   9&  16.1&  20.4&  2.8 &  1.00 \\ 
 ~308    &03:45:54.9 $-$36:21:30 & 1484  &  25 &   16.3&   7&  13.8&  22.3&  7.5 &  1.00 \\ 
 ~310    &03:46:13.8 $-$36:41:48 & 1373  &  13 & $-$1.0&$-$1&  14.0&  20.5&  7.9 &  1.00 \\ 
 ~312    &03:46:19.0 $-$34:56:36 & 1898  &  22 &   22.5&   5&  13.5&   ---&  --- &  1.00 \\ 
 ~316    &03:47:01.4 $-$36:26:15 & 1546  & 105 & $-$0.4&$-$5&  16.9&  23.0&  6.6 &  1.00 \\ 
 ~315    &03:47:04.7 $-$33:42:34 & 1071  &  15 & $-$0.1&   2&  12.6&   ---&  --- &  1.00 \\ 
 ~319    &03:47:16.1 $-$32:18:08 & 1445  &  61 & $-$0.1&$-$5&  17.1&  23.1&  6.6 &  1.00 \\ 
 ~322    &03:47:33.2 $-$38:34:54 &  978  &  15 &   54.4&   7&  12.5&   ---&  --- &  1.00 \\ 
 ~324    &03:47:52.8 $-$36:28:18 & 1493  &  44 & $-$1.1&$-$5&  16.7&  22.4&  5.4 &  1.00 \\ 
 B~2144  &03:49:53.3 $-$32:15:33 & 1184  &  25 &   65.1&$-$4&  17.1&  21.6&  3.2 &  1.00 \\ 
 ~335    &03:50:36.8 $-$35:54:34 & 1367  &  24 & $-$0.1&$-$4&  15.2&  20.5&  4.6 &  1.00 \\ 
 ~336    &03:50:52.6 $-$35:10:19 & 1956  &  67 &    0.4&$-$5&  18.0&  22.7&  3.4 &  1.00 \\ 
 ~338    &03:52:01.1 $-$33:28:07 & 1562  &  16 &   32.9&$-$1&  14.3&   ---&  --- &  1.00 \\ 

\hline
\end{tabular}

Note: (1) P1 is the probability that the galaxy lies in the
Fornax-main cluster and not the Fornax-SW subcluster.}
\end{table*}

\subsection{Membership Status}

\begin{figure}
\psfig{file=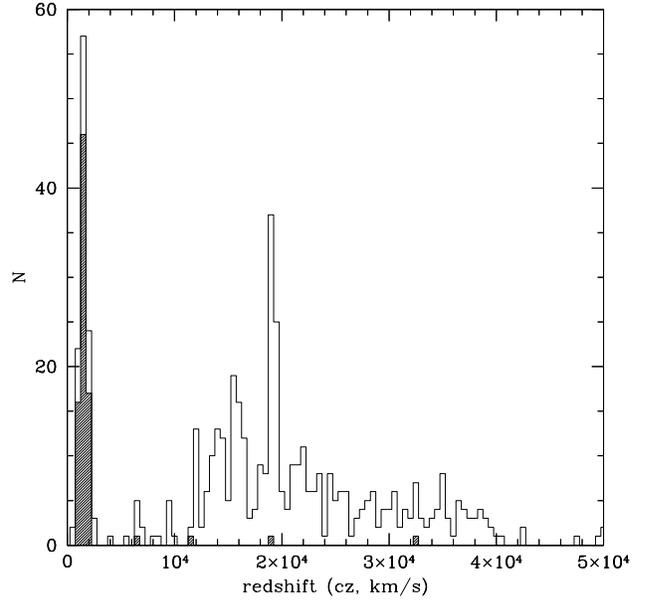,angle=0,width=\one_wide}
\centering
\caption{Distribution of the \nredg\   galaxy velocities measured for this work
(unshaded histogram) compared to the 84 measurements for the same
galaxies in the FCC (shaded histogram).}
\label{fig_redshifts}
\end{figure}

The full impact of our new measurements is demonstrated in
Fig.~\ref{fig_redshifts} which gives the distribution of our \nredg\
measured redshifts compared to the 84 redshifts available in the FCC
for the same galaxies. The majority of our spectra are of background
galaxies as expected since our aim was to search for new cluster
members among galaxies classified as ``background'' in the FCC. The
membership classifications of the objects we observed are compared to
the FCC classifications in Table~\ref{tab_membership}. As in Paper~I,
we detect a clear void behind the Fornax cluster with no galaxies
having redshifts between 2320 and 4180\kms, so we define any galaxy
with a redshift less than 3000\kms\  as a member of the cluster.

\begin{table}
\caption{Comparison of new spectroscopic classification with FCC
membership classes\label{tab_membership}}
\begin{tabular}{lrrrr}
\hline
FLAIR-II classification: & star & member & background & total \\
\hline
FCC member     & 2 & 99 &   3 & 104 \\
FCC background & 5 &  9 & 398 & 412 \\
not in FCC     & 3 & -- &   7 &  10 \\
total          &10 &108 & 408 & 526  \\
\hline
\end{tabular}
\end{table}

We find that the FCC membership classifications were very reliable:
out of \nredg\  galaxies measured, only 9 were members misclassified as
background galaxies in the FCC and only 3 were background galaxies
misclassified as members. These 12 objects with new membership
classifications are listed in Table~\ref{tab_changed}. Note that one
additional new cluster member (FCC B1241) was reported in
Paper~I. Although this galaxy was in our FLAIR-II sample, it was not
observed successfully with FLAIR-II (the identification was from 2dF
observations), so shall not be discussed in this current paper.

\begin{table}
\caption{Galaxies listed in the FCC with changed membership
classifications. 
\label{tab_changed}}
\begin{tabular}{crrrrl}
\hline
 RA (J2000) Dec         & \bj\  &$cz$&$\Delta cz$&FCC&type \\
                        &      &\multicolumn{2}{r}{(\kms)}& \\  
                                  \hline
 03:23:31.7  $-$34:36:34&  15.2 & 13222&   14&    24& BCD    \\        
 03:27:33.8  $-$35:43:04&  17.5 &   723&   79&  B470& S/Im   \\        
 03:31:32.5  $-$38:03:43&  16.5 &  1676&   31&  B729& (d)S0  \\        
 03:33:43.4  $-$35:51:33&  17.9 & 15483&   60&   123& ImV    \\        
 03:33:56.2  $-$34:33:43&  17.4 &  2254&   56&  B904& (d)E   \\        
 03:33:57.2  $-$34:36:43&  17.7 &  1242&   23&  B905& ?      \\        
 03:35:20.4  $-$32:36:08&  13.7 &  1256&   36& B1005& S0     \\        
 03:36:49.7  $-$33:27:39&  17.8 &  1734&  137& B1108& (d)S0  \\
%03:38:16.7  $-$35:30:28&       &  1997&     & B1241& ***    \\
 03:39:55.0  $-$33:03:09&  17.3 &   745&   21& B1379& dE     \\        
 03:41:59.5  $-$35:20:53&  17.7 &  1642&   52& B1554& E      \\        
 03:46:18.2  $-$33:45:48&  17.9 & 47658&   60&   311& dS0    \\        
 03:49:53.3  $-$32:15:33&  17.1 &  1184&   25& B2144& (d)E   \\
\hline
\end{tabular}

Note: the final column (``type'') gives the FCC morphological
classification.
\end{table}		    			

The properties of the new cluster members are discussed in detail
below. Six of these were first presented in Paper~I; our present
analysis has produced an additional three new cluster members. We also
found three background galaxies that were classified as members in the
FCC; these include two of the FCC candidate late-type dwarf galaxies
(FCC 24 and FCC 123). These both had emission line spectra confirming
the late-type classification if not the membership. We inspected
images of all the re-classified objects by eye on the \bj\  survey
plates and there was no apparent reason why they should not have been
given the FCC classifications. This only goes to emphasise the
limitations of the morphological classification.

More interesting, our sample of compact galaxies drawn from the APM
data included 10 objects not in the FCC: three of these were objects
dominated by the light from foreground stars (Table~\ref{tab_star}),
but 7 are background galaxies listed in Table~\ref{tab_not}.  These
were not included in the FCC because of their very compact appearance;
the FCC is diameter-limited at 17\asec, while these objects have scale
lengths of 2\asec\  or less.  Only one has been previously measured
(APMBGC 359+114+047) as part of the Stromlo/APM galaxy redshift survey
\citep{lov1996}; the others we give new names in Table~\ref{tab_not}
according to the IAU-approved convention for our larger Fornax Cluster
Spectroscopic Survey \citep{dri2000}.

\begin{table}
\caption{Background galaxies not listed in the FCC\label{tab_not}}	
\begin{tabular}{lcrrr}
\hline
Name & RA (J2000) Dec            & $cz$&$\Delta cz$ \\
      &                     & \multicolumn{2}{r}{(\kms)} \\  
                             \hline
FCSS J032700.2$-$354854  & 03:27:00.2  $-$35:48:55&   25548&   44 \\
FCSS J033154.8$-$343045  & 03:31:54.9  $-$34:30:45&   21972&   33 \\
FCSS J033744.9$-$344834  & 03:37:45.0  $-$34:48:34&   26232&  133 \\
FCSS J033757.0$-$372542  & 03:37:57.1  $-$37:25:43&   16923&   31 \\
FCSS J034111.5$-$364550  & 03:41:11.5  $-$36:45:50&   31302&   36 \\
APMBGC 359+114+047       & 03:51:17.1  $-$35:45:49&   18102&   51 \\
FCSS J035159.5$-$361420  & 03:51:59.5  $-$36:14:20&   14505&   34 \\
\hline
\end{tabular}             	
\end{table}

We can use the statistical samples defined in Fig.~\ref{fig_morph} to
put limits on the total number of compact new cluster members. In the
upper region in the Figure (above the dashed line) we found 7 new
members having observed 327 galaxies not listed as members in the FCC;
the probability of getting a new member is therefore
$p\approx(7/327)=0.02$. Using this as a binomial probability the
expected number of new members among the remaining 141 galaxies we
didn't observe in this region is therefore equal to 3 and the 98 per
cent upper limit on the additional new members is equal to 7. There is
clearly no large population of new members still to be found in this
magnitude range.

\subsection{Spectroscopic Classification}

Our spectroscopic data not only allow us to confirm the cluster
membership of the Fornax galaxies, but they also give us the
opportunity to revise the classifications of individual cluster
members on the basis of measured spectral indicators of star
formation. The FCC galaxies were given morphological classifications
in the extended Hubble system following the work by Sandage (see
FCC). For the purpose of our current analysis we have converted these
classifications to a one-dimensional sequence of ``revised
morphological types'' or ``t-types'' following \citet[RC3
hereafter]{rc3}. The correspondences between the FCC types and the
t-types we have used are given in Table~\ref{tab_ttypes} which is
based on Table~2 of the RC3 galaxy catalogue except for the early type
dwarfs which are not defined in that table. We assigned dE and dS0
galaxies a t-type of $-$5 based on the classification of some dwarf
galaxies in the Local Group. The one cluster member classified ``?''
we assigned a value of 10 (Im).

\begin{table}
\caption{Correspondence between new galaxy ``t-types'' adopted and FCC
classifications
\label{tab_ttypes}}
\begin{tabular}{rrrr}
\hline
t-type & FCC type & t-type & FCC type \\
\hline
-5 & dS0, dE     & 5 & Sc, SBc     \\
-4 & E           & 6 & Scd, SBcd   \\
-1 & S0, SB0     & 7 & Sd, SBd     \\
 0 & S0/a, S, SB & 8 & Sdm         \\
 1 & Sa, SBa, RSa& 9 & Sm, SBm     \\
 2 & Sab, SBab   &10 & Im, BCD, ?  \\
 3 & Sb, SBb     &15 & other       \\
 4 & Sbc, SBbc   &20 & unclassified\\
\hline
\end{tabular}
\end{table}

\begin{figure}
\psfig{file=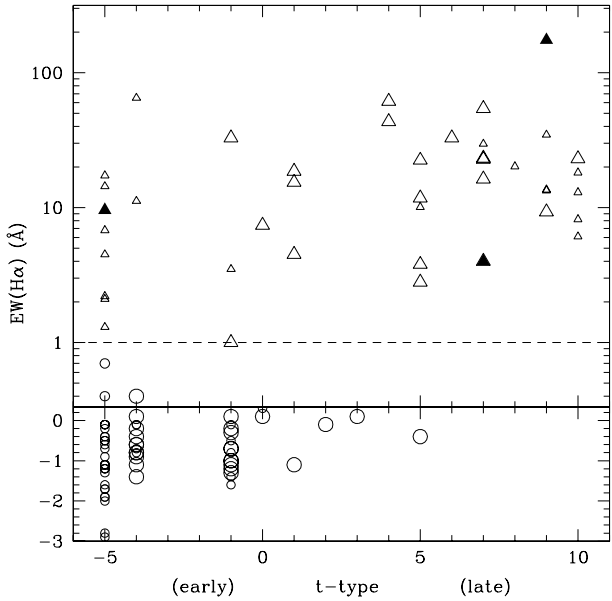,angle=0,width=\one_wide}
%\centering
\caption{Equivalent widths of \ha\ emission as a function of
morphological t-types of 108 Fornax cluster galaxies. The equivalent
widths are measured from our FLAIR-II spectra with positive values
indicating emission. The t-types are derived from the FCC
morphological classifications as discussed in the text. Galaxies with
widths greater than our $3\sigma$ detection limit of 1\AA\ (indicated
by the dashed line) are plotted as triangles and the rest as circles
with the larger symbols in each case for the giant galaxies.  The
three galaxies with possible BCD classifications are indicated by
filled triangles. The scale is logarithmic for widths greater than our
$1\sigma$ limit of 0.3\AA\ (indicated by the horizontal axis) and
linear for smaller values.
\label{fig07ew}}
\end{figure}

The one line detectable in all our spectra is the \Ha\  emission
line. We measured \Ha\  equivalent widths interactively using the
\iraf\  task {\em splot}, defining the continuum level from regions
both sides of the \ha\  line, avoiding the \nii\  lines.  In
Fig.~\ref{fig07ew} we plot the equivalent widths of \Ha\  as a function
of the newly assigned galaxy t-types; the \Ha\  equivalent widths are
listed in Table~\ref{tab_cat}.  There is a general correlation: most
of the late-classified galaxies show emission lines and the dE/dS0
galaxies mostly have low equivalent widths of \Ha. However several of
the early type galaxies do have significant emission in \Ha, notably
the new cluster member FCCB 2144 with an equivalent width of 65\AA\
that was previously classified as ``E or dE(\m32)''. Our 3-sigma
detection for \Ha\  emission corresponds to an equivalent width of
1.0\AA. We classify the 42 galaxies with \Ha\  emission above this
limit as late types and the remainder (66) as early types.

The new spectroscopic classifications reveal a much higher incidence
of star-forming dwarf galaxies in the cluster than implied by the
original morphological classifications, as summarised in
Table~\ref{tab_spectro}. This trend is discussed further in
Section~\ref{sec_newsf} where we also consider the effect of any bias
on the new classifications caused by the limited aperture size of our
observations. It is clear that reliable morphological classifications
would need higher resolution and/or dynamic range material than a wide
field photographic plate, i.e.\ the FCC classifications were
influenced by the data characteristics as noted by Ferguson at the
time.

\begin{table}
\caption{Comparison of new spectroscopic classifications with FCC
morphological types. The dwarf sample is defined as galaxies with
$\bj\geq 15.0$.\label{tab_spectro}}
\begin{tabular}{llrrr}
\hline
Sample & FCC type & $N$  & \multicolumn{2}{c}{Spectroscopic type}\\
       &           &  & absorption & emission  \\
\hline
dwarf  & early  &  50   & 39 & 11 \\
dwarf  & late   &  12   &  1 & 11 \\
giant  & early  &  23   & 21 &  2 \\
giant  & late   &  23   &  5 & 18 \\
total  & --     & 108   & 66 & 42 \\
\hline
\end{tabular}
\end{table}

%%%%%%%%%%%%%%%%%%%%%%%%%%%%%%%%%%%%%%%%%%%%%%%%%%%%%%%%%%%%%%%%%%%%%%%
%%%%%%%%%%%%%%%%%%%%%%%%%%%%%%%%%%%%%%%%%%%%%%%%%%%%%%%%%%%%%%%
\section{Cluster Structure}
\label{sec_structure}

The 108 confirmed cluster galaxies we have measured constitute a very
useful sample to examine the structure of the cluster as they were
uniformly selected over a very large field. This is also the first
spectroscopic sample of cluster galaxies to contain a significant
fraction of dwarf galaxies.  For the purposes of this paper we define
dwarf galaxies to be those with $\bj\geq15$ ($\Mb\geq-16.5$) in our
sample. This is slightly brighter than we have used elsewhere
($\bj\geq15.5$ in Paper~III) so as to give a larger sample (62 dwarfs)
for our analysis, but there is a natural break in the luminosity
distribution at this limit and it matches the luminosity of the
brightest local group dwarfs \citep{mat1998}.  Our analysis of the
galaxy velocities has revealed the first significant evidence for
substructure in the Fornax Cluster. This is presented fully in
Paper~III, so we only summarise it in this section.

\subsection{Cluster Membership  }
\label{sec_newbies}

\begin{figure*}
\vspace{-2.2cm}
\hbox{
\hspace{-1.8cm}
\psfig{file=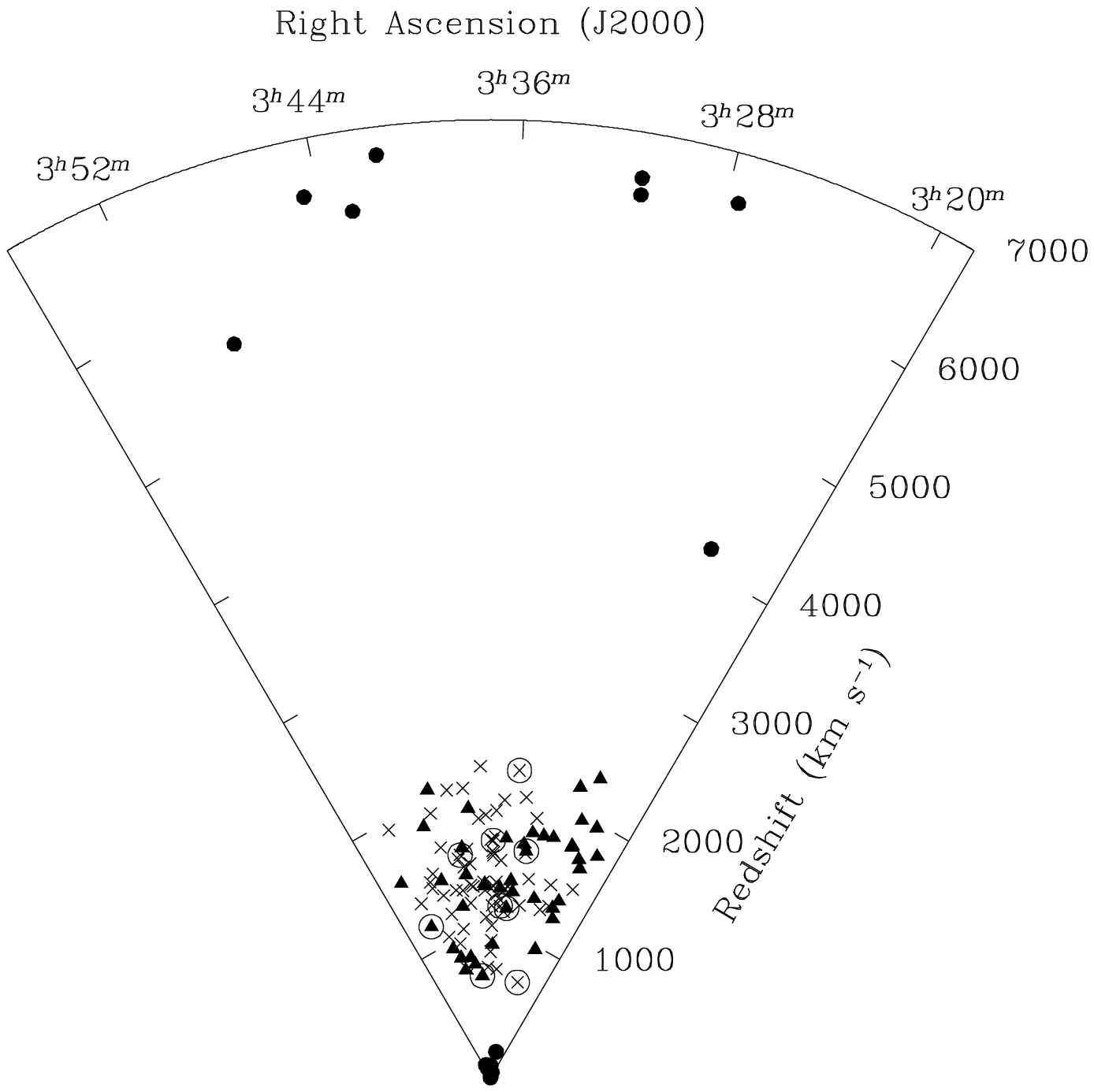,angle=0,width=12cm}
\hspace{-2.9cm}
\psfig{file=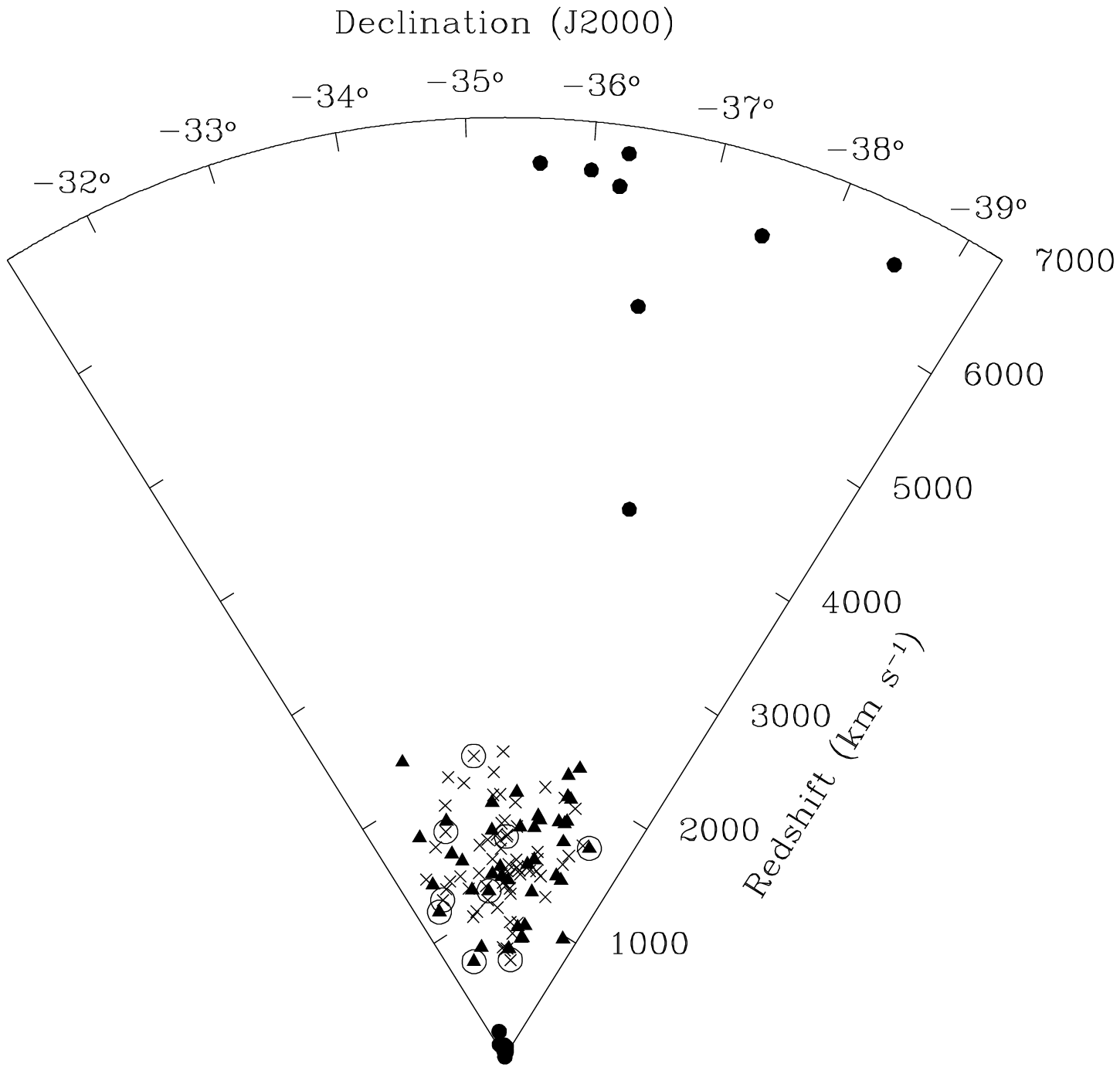,angle=0,width=12cm}
}
\vspace{-1.0cm}
\centering
\caption{Distribution of cluster galaxies. Cluster members are indicated
as early types (crosses), late-types (triangles) and new members
(additional circles). Solid circles indicate foreground
stars and background galaxies.
}
\label{fig_ccone}
\end{figure*}

The overall distribution of the cluster members is shown in
Fig.~\ref{fig_ccone} as cone diagrams in Right Ascension and
Declination. The Figure shows all objects measured between velocities
of zero and 7~000\kms, to emphasise the clear demarcation between the
cluster and foreground stars on the one hand and background galaxies
on the other. The Figure also differentiates between early- and
late-type cluster members as discussed in the next section. The
locations of the nine new cluster members are indicated by circles;
their positions are not significantly different to those of the
previously classified cluster members.

Our new measurements represent a large increase in the number of
confirmed dwarf galaxies in the Fornax Cluster, from 26 with redshifts
in the FCC to 62 in our sample. However as only 9 of our sample were
not already classified as cluster members in the FCC, this does not
represent a significant change in the cluster luminosity function. At
fainter levels ($\bj>18$, $M_B>-13.5$) the galaxy counts in the FCC
tail off rapidly so it is presumably incomplete.

\subsection{Substructure and Dynamics}
\label{sec_velocity}

\begin{figure}
\psfig{file=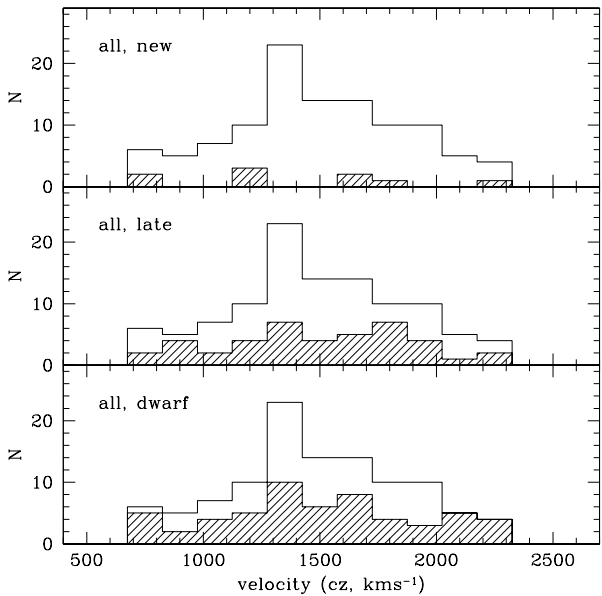,angle=0,width=\one_wide}
\centering
\caption{Histograms of Fornax Cluster galaxy velocities. In each panel
the unshaded histogram shows the full sample of 108 galaxies. The
shaded histograms show various subsets: the nine new cluster members
(top panel), the 42 late-type galaxies (middle) and the 56 dwarf
galaxies (bottom panel).
\label{fig06_histoz}}
\end{figure}

We present histograms of the velocity distributions of cluster
galaxies and various subsamples in Fig.~\ref{fig06_histoz}.  The total
cluster sample has a marginally non-Gaussian velocity distribution at
the 91 per cent\ confidence level using the W-test
\citep{roy1982}. The mean velocity is $1493\pm36\kms$ and the velocity
dispersion is $374\pm26\kms$.

The other subsamples (late- and early-types; dwarfs and giants as
defined in Section~\ref{sec_obs}) are all consistent with Gaussian
distributions. There is however an indication in
Fig.~\ref{fig06_histoz} of differences in the velocity dispersions of
the subsamples. The velocity dispersion of the dwarfs ($409\pm37\kms$)
is larger than that of the giants ($324\pm34\kms$) at the 90 per cent\
confidence level as measured by the F-test \citep{pre1992}. The larger
velocity dispersion of the dwarfs, combined with their more extended
spatial distribution (Section~\ref{sec_spatial}) suggests that they
are infalling whereas the giants form a virialised population (see
Paper~III). The dispersions of the late-type ($405\pm45\kms$) and
early-type samples ($356\pm31\kms$) are not significantly different.

\begin{figure}
\psfig{file=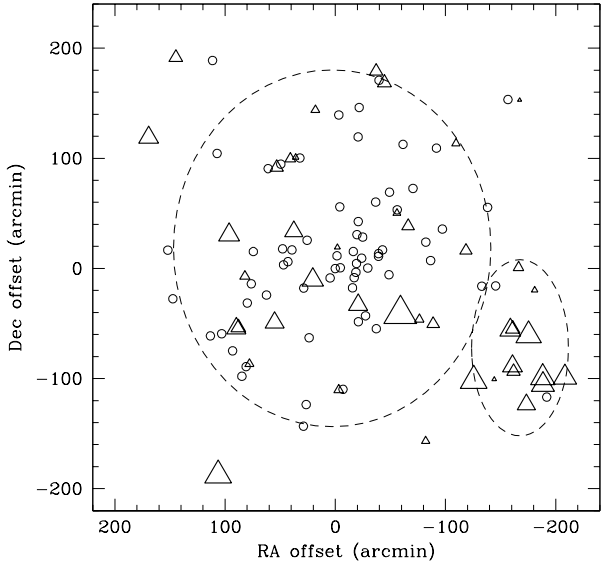,angle=0,width=\one_wide}
\centering
\caption{Projected spatial distribution of Fornax cluster galaxies.
Galaxies spectroscopically classified as early type are plotted as
circles and late types as triangles. The symbol sizes of the late-type
galaxies are scaled by the logarithm of their star formation rates
(ranging from $10^{-3}$ to 6 \moy) derived from their magnitudes and
\ha\  equivalent widths as described in Section~\ref{ssec_evolution}.
The dashed ellipses show the 2-sigma limits of the two subclusters.
\label{figsky}}
\end{figure}

In Paper~III we describe in detail the evidence for substructure in
this galaxy distribution. We used the KMM mixture modelling algorithm
as described by \citet{col1996} to identify a robust partition of the
cluster into a 92-member main cluster centred close to NGC~1399 with
$\overline{cz}=1478\kms$ and $\sigma_{cz}=370\kms$ and a 16-member
subcluster centred about 3 degrees to the South West with
$\overline{cz}=1583\kms$ and $\sigma_{cz}=377\kms$. The partition is
indicated in Table~\ref{tab_cat} in the form of a probability for each
cluster member that it is a member of the main cluster.  Note that the
subcluster does not contain the giant barred spiral NGC 1365 which is
closer to the centre of the field, but it does contain the active
radio galaxy Fornax~A (NGC 1316) as well as a high concentration of
other late-type galaxies. The two-sigma limits of the cluster and
subcluster are shown in Fig.~\ref{figsky}. A simple two-body dynamical
model allows for solutions with the subcluster either in front of the
main cluster (infalling) and behind the main cluster (moving
away) but the infalling solution is slightly more probable.  We use these
dynamical data in Paper~III to make new estimates of the cluster mass
using both virial mass estimators and the velocity amplitude method of
\citet{dia1999}. The cluster mass within a projected radius of
1.4\Mpc\  is $(7\pm2)\times10^{13}\mo$ corresponding to a mass-to-light
ratio of $300\pm 100 \mo/\lo$.

\subsection{Spatial Distribution}
\label{sec_spatial}

\begin{figure}
\psfig{file=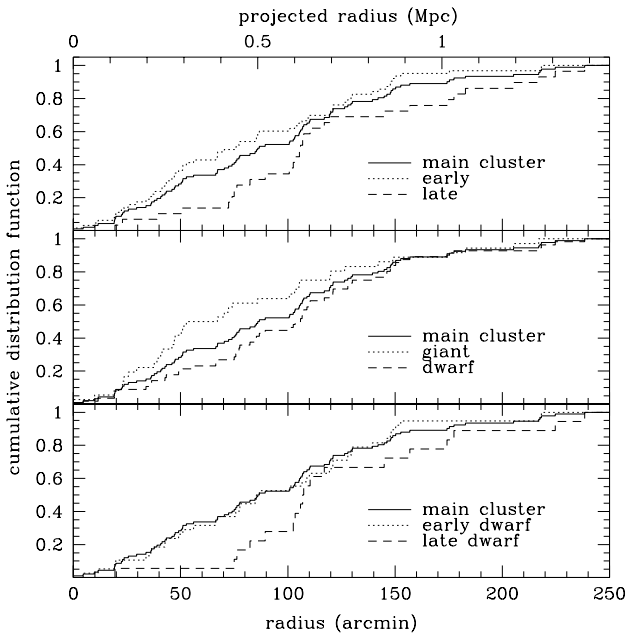,angle=0,width=\one_wide}
\centering
\caption{Cumulative radial distributions of various galaxy samples in
the 92-member main part of the Fornax Cluster. In each plot the total
distribution is shown as a solid line. These are compared to the early
(dotted) and late-type (dashed) subsamples (upper panel), the dwarf
(dotted) and giant (dashed) subsamples (middle panel) and the dwarf
early (dotted) and dwarf late-type (dashed) subsamples (lower panel).}
\label{fig_rad}
\end{figure}

The projected galaxy distribution in Fig.~\ref{figsky} shows evidence
that the late-type galaxies are more widely distributed than the early
types. This is confirmed in Fig.~\ref{fig_rad}, a plot of the
normalised cumulative radial distributions for different galaxy
subsamples. Note that for this radial analysis we restrict our
analysis to the 92 members of the main cluster. Even after the removal
of the SW subcluster which is dominated by late-type galaxies, the
late-type galaxy subsample is significantly more extended than the
rest of the cluster galaxies. The two distributions differ at the 99
per cent\ confidence level measured by the KS test. The dwarf galaxies
are also more extended than the giant galaxies (middle panel of the
figure) at the 99 per cent\ confidence level. Restricting the analysis
to the dwarf galaxies alone (lower panel of the figure) the fraction
of late-type dwarfs also decreases towards the cluster centre, but
given the smaller sample (56 dwarfs) the difference is only at the 85
per cent confidence level.

%%%%%%%%%%%%%%%%%%%%%%%%%%%%%%%%%%%%%%%%%%%%%%%%%%%%%%%%%%%%%%%%%%%%%%%
%%%%%%%%%%%%%%%%%%%%%%%%%%%%%%%%%%%%%%%%%%%%%%%%%%%%%%%%%%%%%%%%%%%%%%%

\section{Star Formation in Dwarf Galaxies}
\label{sec_starf}

\subsection{New star forming galaxies}
\label{sec_newsf}

As described above, we reclassified the measured Fornax Cluster
galaxies as late or early types from their spectroscopic properties,
independent of their morphologies. This spectral reclassification is
based on our FLAIR-II data, i.e.\  spectra taken through 6.7 arc second
fibres centred on the individual galaxies. We therefore need to
consider how representative the spectra are of the whole galaxies,
especially as star formation is often centrally-concentrated in dwarf
galaxies. \citet{gal1989} obtained \ha\  imaging and spectroscopy for a
sample of about 30 Virgo Cluster dwarf irregulars. They found that the
\ha\  emission was significantly centrally-concentrated in 30\% of
their sample. In the remaining galaxies the distribution of \ha\
emission followed that of the red continuum light, in which case central
measurements of \ha\  equivalent width would
be representative of the whole galaxy. A spectroscopic analysis of the
nuclei of Fornax dwarf galaxies \citep{hel1994} found evidence for
young stellar population in some of them. Our sample includes two of
the same galaxies (FCC 207 and 261) in which we detected weak \ha\
emission. \cite{hel1994} remarked that FCC~207 is blue in U-B colour
indicating a young population although their spectra did not extend
far enough into the red to allow them to detect the \ha\  emission.

Ideally we would avoid any bias by classifying galaxies as a function
of their {\em total}  \ha\ emission, normalised in some way by the galaxy size
or mass. The measurement of \ha\ equivalent width (EW) is one example
of this normalisation.  The only reliable way to measure the total
\ha\ emission is with large-aperture spectroscopy or narrow-band \ha\
imaging. This is not available for our sample, although we plan to
obtain \ha\ images of these galaxies in the future. If the \ha\
emission were centrally concentrated then our EW measurements would
over-estimate the average values for each galaxy. We can estimate a
lower limit for the total EW by considering an extreme case where all
the emission is in the central region. The total EW is then just our
measured value multiplied by the fraction of the total galaxy
(continuum) light sampled by the aperture of the spectrograph. The
effective aperture of the the FLAIR-II system is a convolution of the
6.7 arc second fibre diameter with image movement during the
observation (due to tracking errors and differential atmospheric
refraction). These errors are difficult to quantify \citep{par1995a},
but the image translation during long exposures is at least as large
as the fibre apertures because we have observed significant flux
variations between different targets between exposures. The effective
aperture is therefore at least 15 arc seconds in diameter. To
calculate the fraction of light observed for each galaxy, we use the
exponential scale lengths already measured from the APM data (shown in
Fig.~\ref{fig_scale}) and an aperture radius of 8 arc seconds. For a
dwarf with scale length of 4 arc seconds this implies we have observed
60\% of the light and that the lower limit of the the total EW is 60\%
of that measured in our spectra.

We estimated lower limits of total EW for all 11 dwarfs which we had
reclassified as late-type: only one of them resulted in a value lower
than our 1\AA\  cutoff: FCC~36 originally classified as ``dE4 pec,N''
in the FCC. So, although our FLAIR-II spectra are biased towards light
from the galaxy cores (around 8 arc seconds or 800\pc\   in radius), the
detections are strong enough that they would have a significant
detection of \ha\  emission when averaged over the whole galaxy. We are
therefore confident in our use of spectral classifications, although
\ha\  imaging will give us more accurate numbers in the future.

Our spectral reclassification results in a higher fraction of
late-type galaxies than implied by morphological classification,
especially for the dwarfs (see Table~\ref{tab_spectro}). Of the 62
dwarfs measured, 12 were morphologically classified as late-type in
the FCC ($t\geq 0$), but we find that twice this number (22) have
detected \ha\  emission.  The fraction of late-type dwarfs in our
sample is therefore 35 per cent compared to 19 per cent using the FCC
classifications. For the total cluster sample in the FCC, the fraction
(Im types compared to dE+dS0+Im) is 13 per cent.  We compared the
properties of the 11 newly identified star-forming dwarfs with the
other cluster galaxies. Their radial and velocity distributions are
not significantly different to the other galaxies, but we note that
their \ha\  equivalent widths are significantly lower at a KS
significance of 94 per cent. As might be expected, it is the dwarfs
with lower star formation rates that were not identified by morphology
in the FCC. The exception to this is the high star formation galaxies
which were previously misclassified as background objects (see
Paper~I).

\begin{figure*}
\psfig{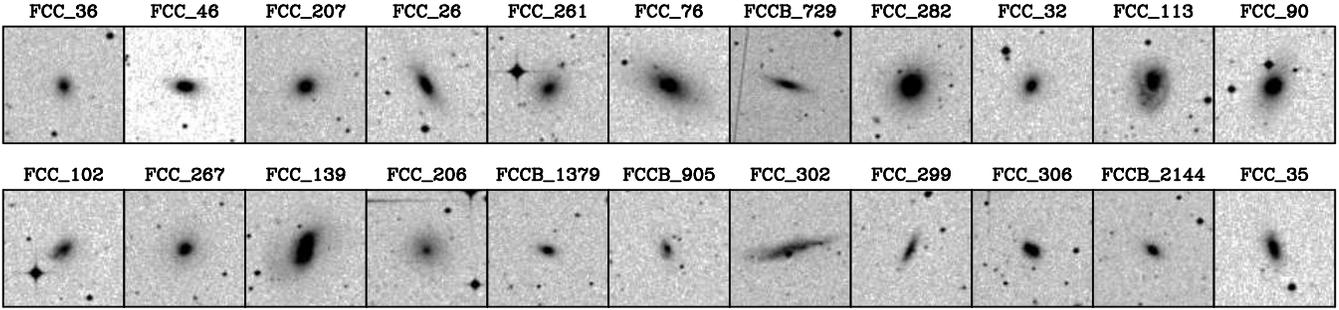}
\centering
\caption{Fornax cluster dwarf galaxies showing significant star
formation plotted in order of increasing \ha\ equivalent widths from
left-to-right and top-to-bottom. The two galaxies previously
classified as BCDs are FCC 32 and FCC 35. These Digitized Sky Survey
(DSS) blue optical images are each about 3 arc minutes across.}
\label{fig_im}
\end{figure*}

We show \bj\  images of all 22 star forming dwarf galaxies in
Fig.~\ref{fig_im} in order of increasing \ha\  equivalent width. The
low equivalent width dwarfs do resemble dE types, but we do not
identify any significant difference in the measured morphological
parameters scale length and magnitude between the old and new
star-forming dwarfs. The new star-forming dwarfs have a slightly
brighter mean surface brightness, but the difference is only
significant at the 84 per cent\  level.

\subsection{Blue compact dwarf galaxies}

Blue compact dwarfs (BCDs) are defined as high surface brightness,
compact, star-forming dwarf galaxies \cite{thu1981}. \citet{fer1989}
identifies a total of 35 candidate BCDs in the FCC, but remarks that
there are very few BCDs in the Fornax Cluster (only five of the
candidates are classified as probable cluster members).

Our spectroscopy of 19 of the candidate BCDs has found that nearly all
of these are background galaxies, with only three actually in the
cluster--FCC~32, FCC~33, and FCC~35. All three have detected \ha\
emission so their BCD classification is confirmed.  Note that FCC~33
is actually too bright to fall in our dwarf sample. Interestingly all
three of these are in fact members of the spiral-rich ``Fornax SW''
subcluster identified in Paper~III and not the main cluster.

\begin{figure}
\psfig{file=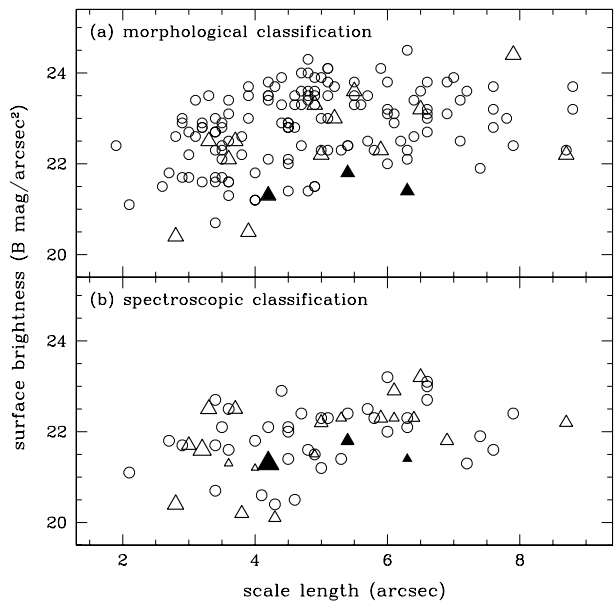,angle=0,width=\one_wide}
\centering
\caption{Diagram of surface brightness against scale length for dwarf
cluster members ($\bj>15.0$). The classifications are shown as early
types (circles) and late-types (triangles). (a) All probable dwarf
cluster members from the FCC using morphological classification; the
three BCDs are shown as filled triangles.  (b) All dwarfs
spectroscopically classified from our FLAIR-II observations. The size
of the triangles is scaled to the logarithm of the \ha\  equivalent
width.}
\label{fig_scale}
\end{figure}

The morphological properties of the FCC-classified BCDs are compared
to the other Fornax dwarf galaxies in Fig.~\ref{fig_scale}(a) which
plots central surface brightness against scale length.  All three BCDs
have relatively high surface brightness (compared to the other FCC
dwarfs) and high \ha\  equivalent widths, consistent with a high rate
of star formation.  However when we include the new cluster members
and spectral classifications in our comparison
(Fig.~\ref{fig_scale}(b)) the FCC-classified BCDs are not so
distinctive as a class; there are several other star-forming dwarfs
with higher surface brightness, although FCC~35 still stands out due
to its exceptionally high \ha\  equivalent width. There is a
significant variation of \ha\  widths with scale lengths: the mean
equivalent width is 37\AA\  for dwarfs with scale lengths less than
4.5\arcsec\  compared to 10\AA\  for larger dwarfs; the distributions
differ at a KS significance of 92 per cent. This suggests that the
star forming region is very compact.

Apart from FCC~35 we cannot identify the FCC-classified BCDs as a
separate class from our measured parameters (scale length, magnitude,
surface brightness or \ha\ equivalent width) in our FLAIR-II
sample. There are other new cluster members that could also be
classified as BCDs, especially those in the low scale length, high
surface brightness region of Fig.~\ref{fig_scale}(b). This agrees with
the conclusions of \citet{mar1999} from a study of nearby blue
``amorphous'' star-forming dwarf galaxies. They found that blue
amorphous galaxies were indistinguishable from blue starburst dwarfs
selected by other means such as BCDs and H~II galaxies. On the basis
of our measurements we can make a stronger statement for the
star-forming dwarfs in the Fornax Cluster: there is no evidence for
two distinct populations of ``compact'' and ``irregular'' star-forming
dwarfs. Instead we find that their distribution of scale sizes is
continuous and is not significantly different to that of the dwarf
ellipticals. The use of ``BCD'' as a separate class is confusing here,
as it implies a dichotomy not seen in our data. We will therefore
avoid any morphological sub-classifications of the late-type dwarfs in
our discussion of their evolution below; we limit ourselves to a
simple classification of the dwarfs into just two classes: early and
late-types based on quantitative \ha\ equivalent width measurements.

\subsection{Evolution of dwarf galaxies}
\label{ssec_evolution}

In previous studies of dwarf galaxies in the Virgo cluster
\citep{dri1991,dri1996} possible evolutionary relationships between
BCDs, dwarf elliptical and dwarf irregular galaxies were examined. It
was concluded that there were very few if any star-forming dwarfs in
Virgo of intermediate size between the BCDs and dwarf irregulars and
in consequence that there was no existing population of star-forming
galaxies that could be progenitors of the dE cluster population
\citep[see also][]{bot1986}. However in the Virgo analysis
\citep{dri1996}, the sample was only complete for very compact
galaxies and relied on morphological classifications for the
intermediate sized dwarfs. Our current sample of Fornax Cluster dwarfs
includes the full range of scale sizes and, in consequence, requires
us to revise the conclusions of \citet{dri1996}.

To see the effects of our improved sample we compare
Fig.~\ref{fig_scale}(b) with the earlier plot of
\citet[][Fig.~4]{dri1991} based on CCD measurements of individual
Virgo dwarfs. From the Virgo sample it was concluded that the star
forming dwarfs--BCDs and irregulars, as defined by morphological
classifications--occupied two extreme regions of the diagram at small
and large scales respectively with no late-type dwarfs at intermediate
scales. In our new data the region of the diagram previously only
occupied by dEs of intermediate scale sizes (5-10 arc seconds) is also
populated with star forming dwarfs: galaxies with the whole range of
scale lengths measured have detected emission lines. On the basis of
this diagram alone we can now revise the conclusion of \citet{dri1996}
that there is no current population of star-forming progenitors of
cluster dwarf ellipticals. As we now find star-forming dwarfs at all
scale lengths, it is much more plausible to model dwarf galaxy
evolution by transformations from late- to early-type dwarfs. To
determine the factors that drive the evolution we must however examine
the distribution of galaxies within the cluster.

The projected distribution of all Fornax galaxies in Fig.~\ref{figsky}
suggests immediately that the star-formation activity in the cluster
is not concentrated in the same regions as the early-type
galaxies. The late-type galaxies are plotted as symbols with sizes
related to their absolute star formation rates. These were calculated
using a relation derived from \citet{ken1992}:
$2.7\times10^{-12}(L_B/L_{B\sun})EW(\ha) \moy$. The Figure shows the
large amount of star formation taking place in the SW subcluster,
3\moy\  out of a total of 11\moy\  for the whole cluster. The giant
galaxies are included in this total star formation rate, so it is
entirely dominated by NGC~1365 (nearer the cluster centre) producing
6\moy.

If we concentrate instead on the dwarf galaxies alone and ignore the
spiral-rich SW subcluster, the radial distributions in
Fig.~\ref{fig_rad}(c) confirm the deficit of star-forming dwarfs at
small projected radii from the centre of the cluster. The fraction of
late-type dwarfs falls from 44 per cent at radii greater than 105
arc minutes to 25 per cent at smaller radii (the radial distributions
differ at a Kolmogorov-Smirnov confidence level of 85 per cent).  This
is consistent with the normal density-morphology relation seen for
more luminous cluster galaxies. The 5 star-forming dwarfs found at
small projected radius from the centre have lower surface brightness
and star formation rates than the other star-forming dwarfs, but the
difference is not statistically significant because of the small
sample. Our data are therefore consistent with the cluster core being
entirely devoid of star-forming dwarfs, the 5 ``central'' dwarfs being
physically located in front of or behind the core.  The lack of any
star-forming dwarfs in the cluster centre provides strong evidence
against a very simple evolutionary model in which all dwarfs undergo
short, sporadic bursts of star formation \citep[e.g.][]{dav1988b}. A
number of processes have been proposed to explain the
density-morphology relation, mainly in the context of the giant galaxy
population of clusters. There is strong evidence from radio
observations that ram pressure stripping can remove the gas from
spirals that pass near the cluster cores \citep{gor1996}, although it
has been argued that this is not such an important process for the
dwarf galaxy population of poor clusters like Virgo and Fornax
\citep{fer1994} because of the lower central gas density. More
recently there has been considerable discussion of dynamical evolution
whereby tidal forces from either the cluster or the larger member
galaxies can result in significant morphological evolution, enough to
convert dwarf irregulars to dwarf ellipticals
\citep{moo1998,may2001}. The second process (``tidal stirring'') was
proposed in the context of galaxies in the Local Group rather than a
rich cluster, but it may also apply in a poor cluster such as Fornax
whose central mass is not much greater than that of the local group.

\begin{figure}
\psfig{file=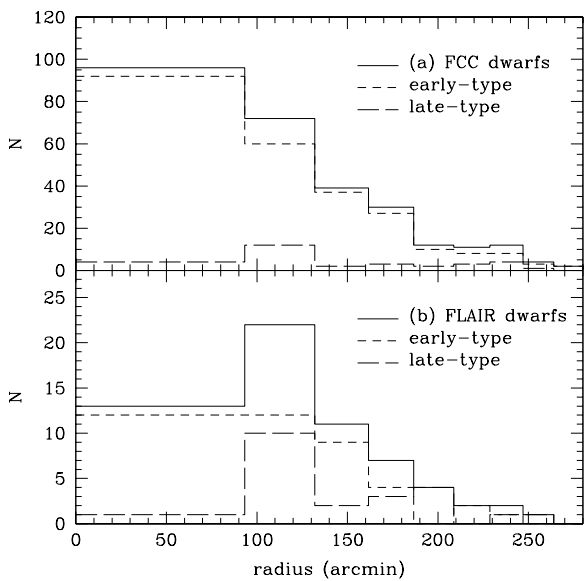,angle=0,width=\one_wide}
\centering
\caption{Radial distribution of dwarf galaxies in the Fornax
Cluster. The plots give the number of galaxies in annuli of equal
area, so the numbers in each bin are directly comparable. In (a) the
dwarfs are separated by their FCC morphological classifications (248
early-type and 31 late-type dwarfs). In (b) the dwarfs are separated
by their spectral classifications from our FLAIR data (40 early-type
and 22 late-type dwarfs). The excess of late-type dwarfs at radii
between 100 and 130 arc minutes is evident in both plots.  }
\label{fig_rdensity}
\end{figure}

The {\em cumulative} radial distribution of late-type dwarfs in
Fig.~\ref{fig_rad}(c) exhibits a sharp jump between projected radii of
100 and 120 arc minutes. This corresponds to an increased density at
that particular radius as is shown in Fig.~\ref{fig_rdensity} which
plots the surface density of the dwarfs as a function of projected
radius. There is a large increase in the number of late-type dwarfs at
projected physical radii around 600\kpc\  which is evident in the FCC
data as well as our smaller FLAIR sample.  This concentration
resembles that of post starburst galaxies at similar radii around rich
clusters reported by \citet{dre1999}. Some process such as interaction
\citep[e.g.][]{mos1993} is triggering star formation as the dwarfs
reach this particular radius. The Fornax dwarfs show significant
evidence of current infall compared to the virialised giant cluster
galaxies (see Paper~III).

Assuming there is no significant gas stripping at this large distance
from the cluster centre, we can estimate the timescale for the current
star formation in some of these dwarfs for which HI measurements are
available \citep{sch2001}. Using their HI masses, we can estimate the
depletion time scales assuming the current rates of star formation
estimated from the \ha\  data and an extra factor of order 2 to allow
for recycling of the gas \citep{ken1994}. The depletion time scales
are $5\times10^9\y$ (FCC~35), $3\times10^{10}\y$ (FCC~113),
$1\times10^{10}\y$ (FCC~139), $5\times10^{11}\y$ (FCC~302) and
$2\times10^{10}\y$ (FCC~306). (The time scales for the giant
star-forming galaxies span a similar range.)  Apart from FCC~35 (a
special case as it is in the active subcluster) these timescales are
all long, of order both the Hubble time and the orbital period of the
cluster at this radius ($3\times10^{10}\y$). If the early-type
galaxies already in the cluster core once had similar properties then
some active process must have removed the gas as they fell into the
cluster; these star formation rates are not high enough to exhaust the
gas in a Hubble time. 

The dwarf galaxies in Fornax appear to trace a complex evolutionary
path. In the cluster core there is virtually no current star
formation, although \cite{hel1994} showed that several of the
nucleated dwarf ellipticals have experienced recent star formation
activity.  This is the classical density-morphology relation seen here
for dwarf galaxies, driven by gas removal and morphological
transformation. In the outer regions of the cluster however, at a
radius of about 600\kpc, there is a concentration of star-forming
dwarfs. This resembles a similar concentration of post starburst
galaxies at similar radii around rich clusters \citep{dre1999}. We
plan to obtain blue spectra of the Fornax dwarfs in order to look for
post starburst galaxies. This combination of enhanced star formation
in the cluster fringes and complete suppression in the cluster core is
very consistent with the results of \cite{has1998} who measured the
spectral properties of a very large sample of galaxies from the Las
Campanas Redshift Survey to determine how star formation rate depends
on the local galaxy density. They found that starburst activity was
enhanced in regions of intermediate density, but that star formation
was suppressed in regions of high density like the centres of galaxy
clusters. 

%%%%%%%%%%%%%%%%%%%%%%%%%%%%%%%%%%%%%%%%%%%%%%%%%%%%%%%%%%%%%%%%%%%%%%%
%%%%%%%%%%%%%%%%%%%%%%%%%%%%%%%%%%%%%%%%%%%%%%%%%%%%%%%%%%%%%%%%%%%%%%%
\section{Compact Dwarf Elliptical (M\thinspace 32-like) candidates }
\label{sec_compacts}

We observed \cdnobs\  of the \m32-like cdE candidates listed in the FCC
(Tables 13 and 3) and successfully determined redshifts for \cdnred\
of these. Only one of these was a cluster member (FCC B2144), but it
had a blue spectrum with emission lines, and therefore is a BCD-like
galaxy, not a normal red dwarf elliptical (details in Paper~I).  The
remaining 75 cdE candidates found to be background objects are
indicated in Table~\ref{tab_back}. 

\begin{figure}
\psfig{file=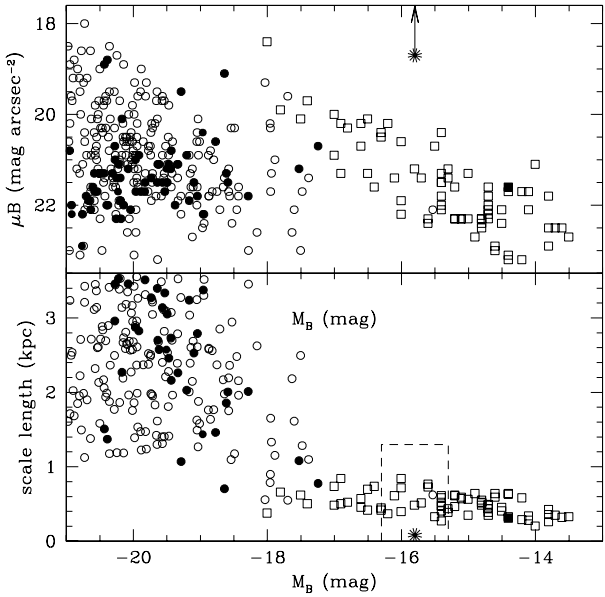,angle=0,width=\one_wide}
\centering
\caption{Structural parameters of candidate cdE (\m32-like) galaxies
with measured redshifts as a function of absolute magnitude. In each
plot background galaxies are plotted as circles and cluster members as
squares; the candidate cdE galaxies (as listed in the FCC) are plotted
as filled symbols.  Upper panel: the distribution of central surface
brightness with the position of \m32\ given by a mean surface
brightness which is a lower limit to the central surface brightness
(see text). Lower panel: the
distribution of physical scale lengths. The position \m32\  would occupy
is indicated by the large asterisk and the region of \m32-like objects
discussed in the text is enclosed by dashed lines. Note that these
structural parameters were calculated from exponential fits to the
photographic images and have not been corrected for seeing or
cosmological fading.}
\label{fig_m32}
\end{figure}

The FCC criteria for cdE candidates included possible background
galaxies, although it was noted that most of these were probably
isolated background ellipticals. We calculated absolute magnitudes and
estimated scale lengths and surface brightnesses for all 75 cdE
candidates in the background sample using a Hubble constant of
75\kmsmpc. These are shown in Fig.~\ref{fig_m32} along with all the
other galaxies we measured for comparison.  Most of the cdE candidates
have redshifts of 10~000\kms\ or more and absolute magnitudes brighter
than \Mb=$-$18 and are giant ellipticals as suggested. Two of the
candidate cdEs were somewhat closer: FCCB~278 at v=7960\kms,
\Mb=$-$17.5 and FCCB~712 at v=6670\kms, \Mb=$-$17.2 with scale lengths
of 1000 and 780\pc\ respectively. These are both much more luminous
and larger than \m32\ which has an absolute magnitude \Mb=$-$15.8
\citep{mat1998} and scale length of 80\pc\ (see below). We therefore
conclude that none of the candidate cdEs listed in the FCC resemble
\m32\ (see locations of filled symbols in Fig.~\ref{fig_m32}).

Following the philosophy of this paper of replacing morphological
classifications by quantitative measures, have included in
Fig.~\ref{fig_m32} the parameters of all the compact-appearing
galaxies we measured in addition to the cdE candidates. We also
indicate the position that \m32\ would occupy in the diagram, although
the values of its surface brightness and scale length parameters are
not directly comparable as they are based on images of much higher
physical resolution than our data. We use the data of \cite{fab1989}
who measure an effective radius of 140\pc\ (corresponding to an
exponential scale length of 80\pc) and a mean surface brightness
within this radius of $\mu B=18.7$ mag arcsec$^{-2}$ which gives a
lower limit for the central surface brightness.

As discussed above, none of the cdE candidates have sizes and
luminosities comparable to \m32\ but there are a number of compact
cluster galaxies in our sample with similar luminosities as indicated
by the dashed line in the Figure. There are 20 galaxies in this
region; nine (including the only background galaxy FCCB~138) have
late-type spectroscopic classifications, but the rest are early-type
dwarf galaxies (most with FCC classifications of dE,N). These are
potentially similar to \m32, but their much larger scale lengths
according to Fig.~\ref{fig_m32} show that they are not like \m32, as
we might expect since they were not classified as cdE candidates in
the FCC.

As a result of our investigation of 75 cdE candidates from the FCC and
our own, larger, sample of compact galaxies, we conclude that there
are no galaxies like \m32\  in the Fornax cluster. A similar search for
\m32-like galaxies in the Leo group was unsuccessful \citep{zie1998},
so this may be a general result.  These observations give some support
to the conclusions of \citet{nie1987} that tidal stripping is not a
realistic formation scenario for such objects.  However we do note
that tidal stripping does appear to be a realistic explanation
\citep{bek2001} for the formation of a less luminous class of compact
galaxy recently found in the Fornax Cluster \citep{dri2000a}

%%%%%%%%%%%%%%%%%%%%%%%%%%%%%%%%%%%%%%%%%%%%%%%%%%%%%%%%%%%%%%%%%%%%%%%
%%%%%%%%%%%%%%%%%%%%%%%%%%%%%%%%%%%%%%%%%%%%%%%%%%%%%%%%%%%%%%%%%%%%%%%
\section{Conclusions}
\label{sec_summary}

In this paper we have made the first detailed study of the Fornax
Cluster in which morphological and membership classifications have
been replaced by quantitative image parameters and spectroscopic
memberships. The image parameters were obtained from complete
catalogues of digitised sky survey plates, allowing for a statistical
approach to our analysis. The spectroscopic data have allowed us to
confirm the membership of many cluster galaxies originally estimated
from visual morphological classification \citep{fer1989}. In our
search for compact cluster galaxies we have found 9 new compact
cluster dwarf galaxies previously classified as background galaxies.

We used our spectroscopic data to estimate star formation rates from
\ha\  emission equivalent widths. The total star formation rate for the
cluster is dominated by the giant spiral NGC~1365, but the remaining
star formation is concentrated in a separate subcluster we have
identified 3 degrees South West of the main cluster, centred on
NGC~1316 (Fornax~A) as described in Paper~III. Among the dwarf
galaxies we find a much higher incidence of detected star formation
than was suggested by the original morphological classifications. In
the 62 dwarf galaxies we measured 35 per cent are star-forming but
only 19 per cent were classified as late types. Our spectral
classification is based on emission from the cores of each galaxy,
but we estimate that a correction to include the whole galaxy would
change at most one of our classifications.

Only three of the BCD candidates were found to be cluster members. Our
data do not support the existence of a well-defined separate class of
BCD-like objects, but rather we detect actively star forming dwarf
galaxies over the whole range of measured sizes. We have also shown
that there is no group of very compact BCDs without corresponding
early-type dwarfs at the same scale sizes. The distribution of scale
sizes is consistent with model of dwarf galaxy evolution involving
morphological transformation from late- to early-type galaxies.  This
is in contrast to earlier work on Virgo Cluster dwarfs \citep{dri1996}
which was not based on spectroscopic classifications.

The fraction of star-forming dwarf galaxies in Fornax drops
significantly towards the cluster centre. This observation alone is
sufficient to rule out simple evolutionary models where the dwarfs
form a single population with all experiencing regular short bursts of
star formation. We estimate long gas depletion timescales for the few
dwarfs with detected H~I emission. This implies that an active process
of gas depletion is needed to transform them to quiescent galaxies
like the central dwarfs.  We conclude that star formation in dwarf
cluster galaxies is a complex process influenced by their environment:
there are less star forming dwarfs at the cluster centre presumably
due to some form of gas stripping. High rates of star formation are
found at the edges of the cluster, perhaps stimulated by tidal
interactions.

Our data also allowed us to look at the candidate compact elliptical
galaxies listed by \citet{fer1989}. We find that none of these have
properties like \m32: they are all background giant ellipticals except
for one cluster member which is less luminous than \m32\  and is also a
spectroscopic late type galaxy. We broadened our search to the larger
sample of compact galaxies we observed and found that, although some
of the brighter cluster dwarfs had similar luminosities to \m32, none
of them had similarly small scale lengths. We therefore confirm our
earlier conclusions \cite{dri1998} that there are no galaxies like \m32\ 
in the Fornax Cluster. We do note however that our 2dF observations
have revealed a population of very compact objects at
the centre of the cluster which resemble low-luminosity
($-13<\Mb<-11$) versions of \m32\   \citep{dri2000a}.

%%%%%%%%%%%%%%%%%%%%%%%%%%%%%%%%%%%%%%%%%%%%%%%%%%%%%%%%%%%%%%%%%%%%%%%
\section*{Acknowledgements}

We are very grateful to all the staff of the UK Schmidt Telescope for
the assistance with our many observing runs. The referee of this paper
made a number of suggestions which have greatly improved the
presentation of this work. We also wish to thank Jon
Davies for a copy of his galaxy catalogue with the APM photometry,
Bryn Jones for discussions of morphological t-types, and Marion
Schmitz (of NED) for checking our galaxy lists prior to publication.
Part of this work was done at the Institute of Geophysics and
Planetary Physics, under the auspices of the U.S. Department of Energy
by Lawrence Livermore National Laboratory under contract
No.~W-7405-Eng-48. This material is based upon work supported by the
National Science Foundation under Grant No.~9970884.

The Digitized Sky Surveys (DSS) were produced at the Space Telescope
Science Institute under U.S. Government grant NAG W-2166. The images
of these surveys are based on photographic data obtained using the
Oschin Schmidt Telescope on Palomar Mountain and the UK Schmidt
Telescope. The plates were processed into the present compressed
digital form with the permission of these institutions.  The UK
Schmidt Telescope was operated by the Royal Observatory Edinburgh,
with funding from the UK Science and Engineering Research Council
(later the UK Particle Physics and Astronomy Research Council), until
1988 June, and thereafter by the Anglo-Australian Observatory. The
blue plates of the southern Sky Atlas and its Equatorial Extension
(together known as the SERC-J), as well as the Equatorial Red (ER),
and the Second Epoch [red] Survey (SES) were all taken with the UK
Schmidt.

%\begin{thebibliography} %\8l %\Y,%\j,%\V,%\p\n

%\end{thebibliography}
%
% add Paper~I and III notes.
%

\appendix
%%%%%%%%%%%%%%%%%%%%%%%%%%%%%%%%%%%%%%%%%%%%%%%%%%%%%%%%%%%%%%%
\section{Background galaxies}
\label{sec_back}

The majority of the galaxies we measured background objects behind the
Fornax Cluster as expected from our compact selection criteria. We
list these 408 background galaxies in Table~\ref{tab_back} below.

We note that there are two background clusters identified behind the
central region of the Fornax Cluster: a poor cluster at $z=0.11$
\citep{hil1999b} and a more distant cluster (``J1556.15BL'') at unknown
redshift \citep{cou1991}. Neither of these are evident in our data,
the first because the members are fainter than our magnitude limit and
the second, presumably, because it is at too high a redshift.

\begin{figure*}
\hbox{
\psfig{file=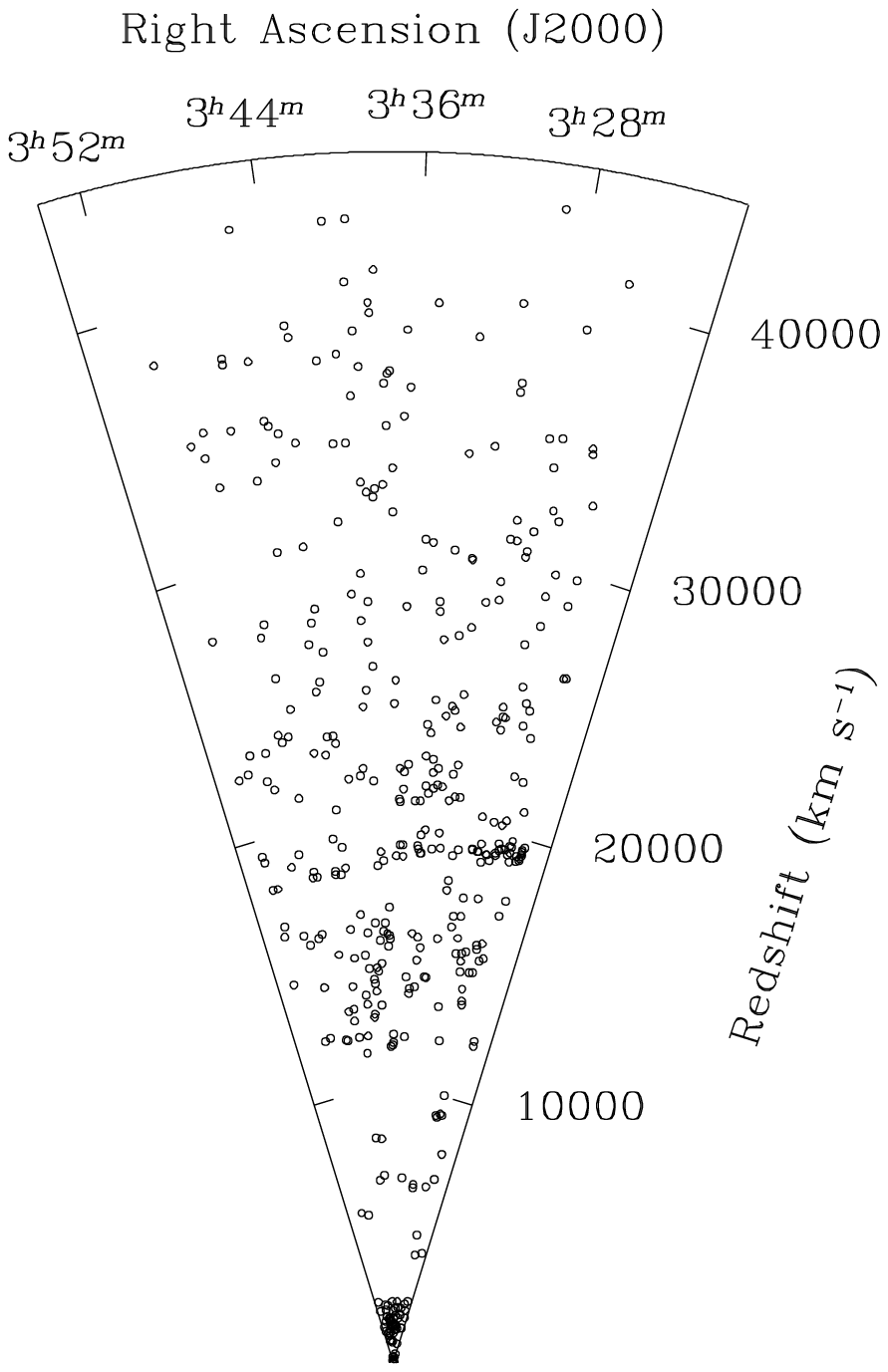,angle=0,width=\one_wide}
\psfig{file=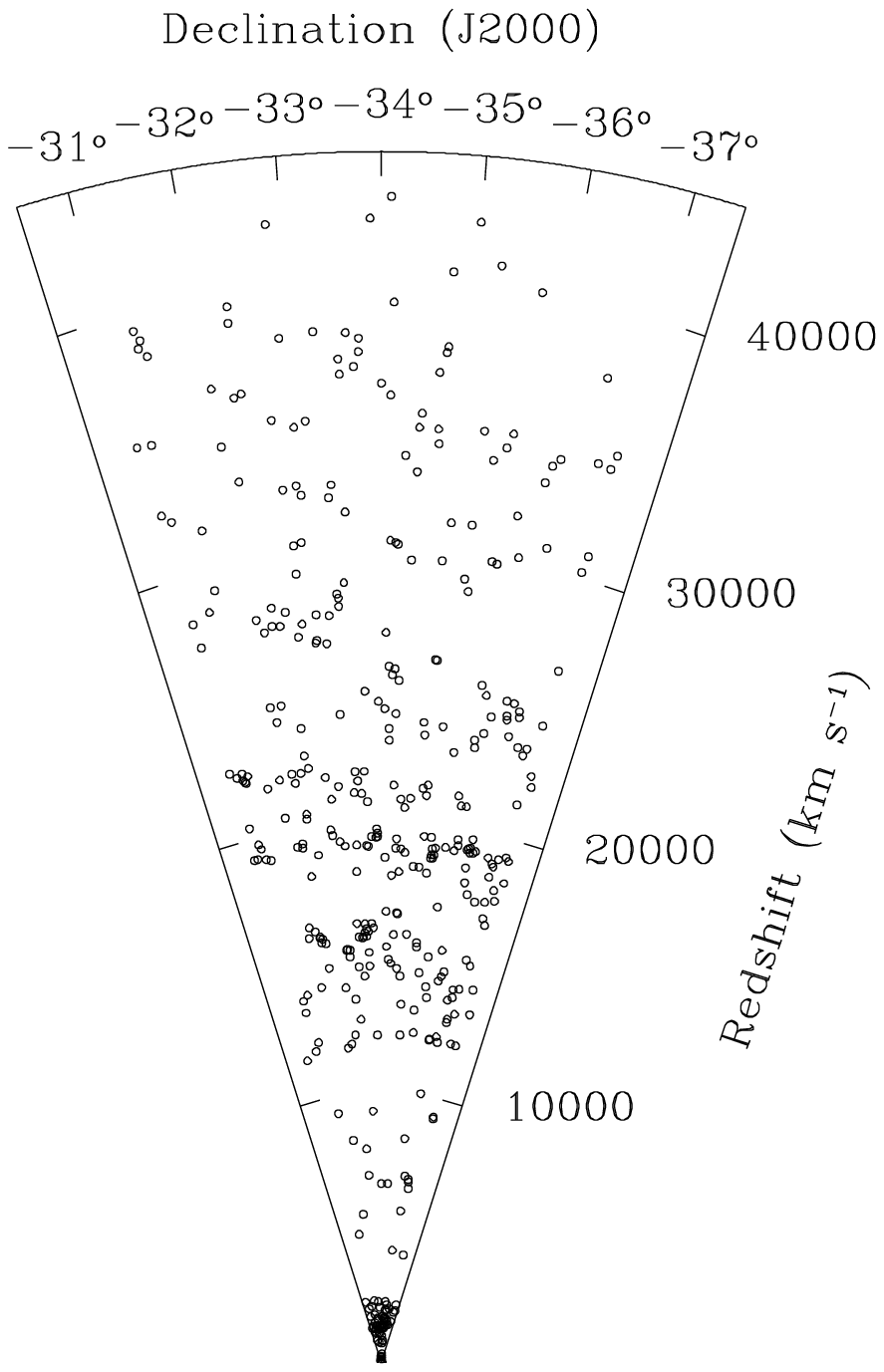,angle=0,width=\one_wide}
}
\centering
\caption{Distribution of the observed galaxies.}
\label{fig05_cone}
\end{figure*}

\begin{table*}
\caption{Catalogue of all background galaxies with measured
redshifts\label{tab_back}}
{\tiny
\begin{tabular}{rrrrrrrrrr}
\hline
FCC&RA (J2000) Dec&cz& $\sigma_{cz}$&cdE & FCC&RA (J2000) Dec&cz&
$\sigma_{cz}$ & cdE  \\
\hline
FCCB~0102  &03:22:26.6  $-$36:15:30  &  25572   &    99   &   &  FCCB~0599  &03:29:23.6  $-$37:21:28  &  12069   &    27  &  \\ 
FCCB~0116  &03:22:42.7  $-$36:26:26  &  19364   &    59   &   &  FCCB~0598  &03:29:25.9  $-$35:08:29  &  13926   &    38  &*  \\ 
FCCB~0119  &03:22:46.6  $-$36:21:12  &  19231   &    43   &   &  FCCB~0600  &03:29:28.1  $-$34:29:12  &  31614   &    73  &  \\ 
FCCB~0126  &03:22:52.9  $-$36:46:03  &  12131   &    45   &   &  FCCB~0605  &03:29:31.0  $-$37:28:31  &  19126   &    43  &  \\ 
FCCB~0127  &03:22:57.8  $-$34:17:16  &  15727   &    21   &   &  FCCB~0616  &03:29:47.9  $-$36:46:57  &  29276   &    77  &  \\ 
FCCB~0138  &03:23:02.4  $-$36:54:09  &   4184   &    34   &   &  FCCB~0624  &03:29:55.4  $-$34:39:33  &  28591   &    97  &  \\ 
FCCB~0153  &03:23:19.0  $-$36:00:22  &  19729   &    85   &   &  FCCB~0631  &03:30:01.4  $-$35:36:26  &  16719   &    20  &  \\ 
FCCB~0162  &03:23:23.7  $-$37:32:10  &  12113   &    46   &   &  FCCB~0638  &03:30:08.2  $-$38:08:58  &  29280   &    93  &  \\ 
FCC~24     &03:23:31.7  $-$34:36:34  &  13222   &    14   &   &  FCCB~0634  &03:30:10.7  $-$34:14:59  &  16333   &    67  &*  \\ 
FCCB~0185  &03:23:37.6  $-$36:05:50  &  19296   &    55   &   &  FCCB~0653  &03:30:24.5  $-$36:22:35  &  19206   &    41  &  \\ 
FCCB~0183  &03:23:38.2  $-$35:06:53  &  26186   &    44   &   &  FCCB~0650  &03:30:25.3  $-$33:38:53  &  36348   &   102  &  \\ 
FCCB~0178  &03:23:38.2  $-$33:27:42  &  20402   &    36   &   &  FCCB~0651  &03:30:25.7  $-$33:58:58  &  36699   &    52  &  \\ 
FCCB~0187  &03:23:38.2  $-$36:53:14  &  12315   &    51   &   &  FCCB~0654  &03:30:27.3  $-$34:40:15  &  19017   &    32  &  \\ 
FCCB~0192  &03:23:39.6  $-$36:54:31  &  19608   &    41   &   &  FCCB~0657  &03:30:30.5  $-$35:22:08  &  13335   &    22  &  \\ 
FCCB~0208  &03:23:51.3  $-$35:05:34  &  26163   &    65   &   &  FCCB~0674  &03:30:41.9  $-$36:55:16  &  14002   &    21  &  \\ 
FCCB~0206  &03:23:52.1  $-$34:05:14  &  19553   &    86   &   &  FCCB~0672  &03:30:44.4  $-$34:38:45  &  28736   &    99  &  \\ 
FCCB~0213  &03:23:56.4  $-$34:54:29  &  19408   &    61   &   &  FCCB~0679  &03:30:45.0  $-$38:11:24  &  12321   &    43  &  \\ 
FCCB~0216  &03:23:58.7  $-$34:19:40  &  19229   &    88   &   &  FCCB~0676  &03:30:49.7  $-$34:38:44  &  28444   &    35  &  \\ 
FCCB~0219  &03:23:59.5  $-$36:03:57  &  19350   &    33   &   &  FCCB~0693  &03:30:57.7  $-$37:19:01  &  18018   &    56  &  \\ 
FCCB~0248  &03:24:24.0  $-$37:20:32  &  24905   &    99   & * &  FCCB~0692  &03:31:00.1  $-$35:28:09  &  18974   &    76  &*  \\ 
FCCB~0245  &03:24:25.6  $-$34:22:26  &  19371   &    42   &   &  FCCB~0691  &03:31:01.2  $-$34:23:28  &  21139   &    61  &  \\ 
FCCB~0249  &03:24:26.5  $-$36:35:27  &  17043   &    37   &   &  FCCB~0702  &03:31:04.6  $-$37:55:29  &  17661   &    24  &  \\ 
FCCB~0255  &03:24:36.6  $-$34:06:14  &  15265   &   101   &   &  FCCB~0703  &03:31:08.6  $-$37:07:09  &  34247   &    45  &  \\ 
FCCB~0252  &03:24:36.8  $-$32:34:13  &  19156   &    58   &   &  FCCB~0707  &03:31:12.6  $-$36:14:26  &   6543   &    53  &  \\ 
FCCB~0258  &03:24:39.3  $-$35:33:53  &  29831   &    39   &   &  FCCB~0712  &03:31:19.6  $-$35:35:01  &   6666   &    47  &*  \\ 
FCCB~0265  &03:24:44.8  $-$36:42:48  &  28839   &    78   &   &  FCCB~0718  &03:31:26.3  $-$35:05:07  &  21130   &    33  &  \\ 
FCCB~0266  &03:24:45.5  $-$36:19:42  &  19634   &    62   &   &  FCCB~0727  &03:31:30.3  $-$37:42:04  &  23734   &    45  &  \\ 
FCCB~0272  &03:24:48.0  $-$36:18:14  &  19421   &    42   &   &  FCCB~0740  &03:31:40.6  $-$37:21:57  &  50131   &    91  &  \\ 
FCCB~0274  &03:24:52.0  $-$37:04:32  &  22077   &    42   &   &  FCCB~0734  &03:31:41.5  $-$34:17:03  &  24942   &    56  &  \\ 
FCCB~0276  &03:24:56.4  $-$34:39:17  &  32670   &    99   & * &  FCCB~0731  &03:31:41.8  $-$32:34:03  &  22336   &    80  &*  \\ 
FCCB~0278  &03:24:58.0  $-$34:47:17  &   7957   &    38   & * &  FCCB~0743  &03:31:48.2  $-$34:23:45  &  15838   &    39  &  \\ 
FCCB~0288  &03:25:12.1  $-$33:05:02  &  19875   &    66   &   &  FCCB~0747  &03:31:49.5  $-$37:04:10  &  18290   &    27  &  \\ 
FCCB~0290  &03:25:16.4  $-$32:28:14  &  19095   &    74   &   &  FCCB~0753  &03:31:51.8  $-$38:14:52  &  12129   &    64  &  \\ 
FCCB~0305  &03:25:23.5  $-$37:15:55  &  16334   &    37   &   &     Name 2  &03:31:54.9  $-$34:30:45  &  21972   &    33  &  \\ 
FCCB~0298  &03:25:23.5  $-$33:03:25  &  19675   &    62   & * &  FCCB~0754  &03:31:55.6  $-$36:21:07  &  29967   &    35  &  \\ 
FCCB~0309  &03:25:30.5  $-$34:52:08  &  15359   &    54   & * &  FCCB~0757  &03:31:59.5  $-$35:24:34  &  20979   &    23  &  \\ 
FCCB~0311  &03:25:34.3  $-$33:52:21  &  14306   &    41   &   &  FCCB~0761  &03:31:59.5  $-$36:25:17  &  30043   &    71  &*  \\ 
FCCB~0315  &03:25:38.9  $-$32:01:07  &  13763   &    30   & * &  FCCB~0774  &03:32:04.1  $-$37:59:58  &  20140   &    59  &  \\ 
FCCB~0331  &03:25:43.3  $-$37:31:26  &  16664   &    39   & * &  FCCB~0777  &03:32:08.2  $-$37:47:21  &  24327   &    43  &  \\ 
FCCB~0332  &03:25:45.0  $-$37:35:60  &  24721   &    64   &   &  FCCB~0786  &03:32:17.0  $-$37:22:42  &  19194   &    46  &*  \\ 
FCCB~0328  &03:25:46.8  $-$33:58:56  &  34526   &    63   &   &  FCCB~0792  &03:32:24.1  $-$37:48:14  &  24476   &    51  &  \\ 
FCCB~0339  &03:25:53.4  $-$35:40:14  &  15902   &    50   &   &  FCCB~0793  &03:32:24.9  $-$38:05:47  &  17712   &    67  &  \\ 
FCCB~0336  &03:25:53.6  $-$33:43:37  &  19490   &    73   &   &  FCCB~0797  &03:32:34.1  $-$33:18:19  &  27130   &    78  &*  \\ 
FCCB~0337  &03:25:53.6  $-$34:19:52  &  15251   &    44   &   &  FCCB~0801  &03:32:35.9  $-$36:33:10  &  11986   &    22  &  \\ 
FCCB~0346  &03:25:59.0  $-$36:01:43  &  19421   &    58   & * &  FCCB~0805  &03:32:49.4  $-$32:40:07  &  19756   &   111  &  \\ 
FCCB~0347  &03:26:02.2  $-$34:06:35  &  19506   &    29   &   &  FCCB~0817  &03:32:51.1  $-$34:51:07  &  21467   &    60  &  \\ 
FCCB~0358  &03:26:08.2  $-$37:43:10  &  24981   &    69   &   &  FCCB~0816  &03:32:51.5  $-$34:29:30  &  14374   &    42  &*  \\ 
FCCB~0359  &03:26:13.0  $-$34:48:40  &  15655   &    78   &   &  FCCB~0815  &03:32:53.3  $-$33:08:22  &  38232   &    37  &*  \\ 
FCCB~0362  &03:26:16.8  $-$33:33:02  &  20554   &    74   &   &  FCCB~0829  &03:33:00.6  $-$35:27:07  &  14369   &    47  &  \\ 
FCCB~0373  &03:26:17.1  $-$37:54:14  &  19297   &    80   &   &  FCCB~0838  &03:33:04.4  $-$37:07:13  &  23859   &    22  &  \\ 
FCCB~0379  &03:26:20.3  $-$36:48:05  &  19222   &    37   &   &  FCCB~0847  &03:33:13.4  $-$35:04:29  &  21510   &    35  &*  \\ 
FCCB~0378  &03:26:22.7  $-$35:02:28  &  15650   &    83   & * &  FCCB~0854  &03:33:14.1  $-$36:34:02  &  30148   &    37  &  \\ 
FCCB~0376  &03:26:23.5  $-$34:05:03  &  19686   &   102   & * &  FCCB~0858  &03:33:19.3  $-$35:20:43  &  21157   &    35  &  \\ 
FCCB~0389  &03:26:29.5  $-$36:10:19  &  31840   &    44   &   &  FCCB~0859  &03:33:19.6  $-$35:40:32  &  19122   &    58  &  \\ 
FCCB~0388  &03:26:33.0  $-$32:40:05  &  22219   &    66   &   &  FCCB~0860  &03:33:24.7  $-$32:38:24  &  22145   &    58  &*  \\ 
FCCB~0391  &03:26:35.1  $-$33:30:21  &  14783   &    42   &   &  FCCB~0871  &03:33:36.2  $-$35:10:03  &  21396   &    93  &*  \\ 
FCCB~0403  &03:26:41.7  $-$34:59:35  &  27896   &    51   & * &  FCCB~0874  &03:33:40.6  $-$33:00:54  &  26930   &   105  &*  \\ 
FCCB~0413  &03:26:48.7  $-$34:15:12  &  26202   &   105   &   &  FCCB~0883  &03:33:41.4  $-$38:21:52  &  20210   &    55  &  \\ 
FCCB~0420  &03:26:51.5  $-$34:22:24  &   9351   &    60   &   &  FCC~123    &03:33:43.4  $-$35:51:33  &  15483   &    60  &  \\ 
FCCB~0422  &03:26:51.5  $-$35:14:37  &  19731   &    29   &   &  FCCB~0882  &03:33:46.9  $-$34:19:44  &  21982   &    57  &*  \\ 
FCCB~0424  &03:26:53.2  $-$34:20:04  &  14742   &    33   &   &  FCCB~0902  &03:33:53.9  $-$35:51:43  &  15463   &    42  &  \\ 
   Name 1  &03:27:00.2  $-$35:48:55  &  25548   &    44   &   &  FCCB~0897  &03:33:56.4  $-$32:39:37  &  22471   &    68  &  \\ 
FCCB~0438  &03:27:00.3  $-$36:35:42  &  13660   &    34   &   &  FCCB~0909  &03:34:01.4  $-$33:55:24  &  24673   &    42  &  \\ 
FCCB~0446  &03:27:09.2  $-$37:09:11  &  19202   &    64   & * &  FCCB~0920  &03:34:03.3  $-$37:19:43  &  23435   &    64  &  \\ 
FCCB~0443  &03:27:09.3  $-$34:47:41  &  13538   &    51   &   &  FCCB~0916  &03:34:06.5  $-$33:54:32  &  27963   &    81  &  \\ 
FCCB~0447  &03:27:09.4  $-$38:19:05  &   9287   &    65   &   &  FCCB~0921  &03:34:07.5  $-$34:32:23  &  21186   &    61  &  \\ 
FCCB~0448  &03:27:11.5  $-$37:27:52  &  17520   &    45   & * &  FCCB~0926  &03:34:09.3  $-$34:05:54  &  19833   &    54  &*  \\ 
FCCB~0441  &03:27:11.5  $-$32:42:16  &  16671   &    34   &   &  FCCB~0925  &03:34:10.0  $-$33:39:14  &  28322   &    53  &  \\ 
FCCB~0445  &03:27:12.9  $-$33:24:10  &  19001   &    71   & * &  FCCB~0931  &03:34:11.5  $-$37:48:50  &  19136   &    39  &  \\ 
FCCB~0453  &03:27:16.7  $-$36:51:57  &  19167   &    60   &   &  FCCB~0939  &03:34:21.7  $-$35:45:09  &  21452   &    35  &  \\ 
FCCB~0450  &03:27:18.0  $-$34:41:47  &  32180   &    65   &   &  FCCB~0938  &03:34:24.8  $-$33:04:30  &  13972   &    81  &  \\ 
FCCB~0454  &03:27:22.6  $-$33:56:40  &   9392   &    45   &   &  FCCB~0943  &03:34:26.0  $-$36:49:09  &  18958   &    50  &  \\ 
FCCB~0459  &03:27:25.2  $-$35:59:49  &   6640   &    32   &   &  FCCB~0946  &03:34:28.3  $-$36:53:26  &  23740   &   123  &  \\ 
FCCB~0458  &03:27:26.8  $-$33:40:32  &  39044   &   102   & * &  FCCB~0950  &03:34:31.6  $-$36:52:20  &  18946   &    79  &  \\ 
FCCB~0466  &03:27:30.8  $-$36:35:02  &  33769   &   106   &   &  FCCB~0948  &03:34:32.5  $-$35:03:55  &  19618   &    21  &  \\ 
FCCB~0469  &03:27:30.9  $-$37:36:28  &   9253   &    39   &   &  FCCB~0951  &03:34:37.7  $-$35:33:30  &  28887   &    63  &*  \\ 
FCCB~0465  &03:27:35.0  $-$33:03:58  &  34898   &    96   & * &  FCCB~0962  &03:34:46.0  $-$36:21:51  &  20894   &    58  &*  \\ 
FCCB~0477  &03:27:36.5  $-$37:27:23  &  18887   &    24   &   &  FCCB~0963  &03:34:48.0  $-$35:33:31  &  19118   &    68  &*  \\ 
FCCB~0475  &03:27:37.7  $-$33:48:24  &  14121   &   190   & * &  FCCB~0960  &03:34:49.2  $-$33:13:22  &  30514   &    89  &  \\ 
FCCB~0471  &03:27:38.0  $-$32:46:02  &  20428   &    62   &   &  FCCB~0968  &03:34:50.8  $-$36:02:03  &  19213   &    40  &  \\ 
FCCB~0482  &03:27:43.7  $-$35:01:49  &  15508   &    86   & * &  FCCB~0986  &03:35:05.9  $-$35:51:31  &  39410   &    95  &*  \\ 
FCCB~0490  &03:27:45.5  $-$37:36:15  &   9331   &    32   &   &  FCCB~0994  &03:35:06.0  $-$37:54:51  &  19073   &    64  &  \\ 
FCCB~0501  &03:27:57.9  $-$36:03:53  &  24336   &    88   &   &  FCCB~0981  &03:35:06.0  $-$33:38:12  &  15970   &    77  &  \\ 
FCCB~0502  &03:27:58.3  $-$35:56:34  &  24294   &    90   &   &  FCCB~0984  &03:35:08.6  $-$32:16:16  &  13908   &    40  &  \\ 
FCCB~0513  &03:28:08.0  $-$35:02:44  &  31301   &    51   &   &  FCCB~0993  &03:35:13.3  $-$32:25:26  &  13726   &    25  &  \\ 
FCCB~0516  &03:28:13.4  $-$34:12:35  &  19550   &    39   &   &  FCCB~1009  &03:35:17.6  $-$36:15:07  &  20881   &    41  &  \\ 
FCCB~0517  &03:28:14.4  $-$34:38:37  &  14735   &    68   & * &  FCCB~1004  &03:35:19.4  $-$33:06:42  &  30604   &    99  &  \\ 
FCCB~0522  &03:28:15.6  $-$36:03:46  &  30321   &    52   &   &  FCCB~1014  &03:35:19.5  $-$36:22:38  &  18888   &    52  &  \\ 
FCCB~0532  &03:28:24.3  $-$34:52:25  &  15107   &    28   &   &  FCCB~1013  &03:35:20.0  $-$35:59:30  &  15564   &    61  &*  \\ 
FCCB~0535  &03:28:27.6  $-$34:55:45  &  15401   &    79   &   &  FCCB~1020  &03:35:29.6  $-$33:13:55  &  29463   &   139  &  \\ 
FCCB~0542  &03:28:38.4  $-$32:49:37  &  19235   &    62   &   &  FCCB~1027  &03:35:34.4  $-$36:18:54  &  12111   &   104  &  \\ 
FCCB~0548  &03:28:43.4  $-$33:59:52  &  20549   &    40   &   &  FCCB~1029  &03:35:39.4  $-$36:17:48  &  14341   &    48  &  \\ 
FCCB~0547  &03:28:44.3  $-$32:44:54  &  19238   &    34   &   &  FCCB~1034  &03:35:47.5  $-$32:58:42  &  38808   &    60  &*  \\ 
FCCB~0556  &03:28:46.5  $-$37:13:34  &  19310   &    42   &   &  FCCB~1056  &03:36:07.1  $-$33:24:02  &  22243   &    30  &  \\ 
FCCB~0561  &03:28:54.0  $-$36:34:47  &  19276   &    65   &   &  FCCB~1081  &03:36:22.3  $-$38:10:19  &  19068   &    61  &  \\ 
FCCB~0567  &03:28:56.5  $-$38:04:18  &  19855   &    52   &   &  FCCB~1083  &03:36:25.1  $-$36:53:30  &  18824   &    45  &  \\ 
FCCB~0566  &03:28:56.7  $-$36:05:05  &  19156   &    39   &   &  FCCB~1087  &03:36:27.5  $-$37:15:43  &  21315   &   106  &  \\ 
FCCB~0570  &03:28:57.5  $-$37:42:46  &  17456   &    73   & * &  FCCB~1090  &03:36:27.7  $-$37:57:25  &  19234   &    44  &  \\ 
FCCB~0576  &03:29:05.0  $-$35:19:34  &  15943   &    76   &   &  FCCB~1079  &03:36:27.9  $-$33:58:28  &  21953   &    96  &  \\ 
FCCB~0582  &03:29:10.0  $-$35:52:35  &  16759   &    40   &   &  FCCB~1084  &03:36:30.2  $-$33:44:01  &  36261   &   105  &*  \\ 
FCCB~0580  &03:29:10.2  $-$34:57:41  &  15360   &    46   &   &  FCCB~1085  &03:36:30.9  $-$33:27:26  &  28100   &    87  &  \\ 
FCCB~0594  &03:29:22.3  $-$37:06:60  &  30911   &   107   & * &  FCCB~1094  &03:36:38.3  $-$34:22:56  &  17777   &    48  &  \\ 

\hline
\end{tabular}}

Notes: the names of the galaxies not listed in the FCC are given in
Table~\ref{tab_not}. The asterisks indicate cdE candidates from the FCC.
\end{table*}

\begin{table*}
\caption{(cont'd) Catalogue of all background galaxies with measured
redshifts}
{\tiny
\begin{tabular}{rrrrrrrrrr}
\hline
FCC&RA (J2000) Dec&cz& $\sigma_{cz}$&cdE & FCC&RA (J2000) Dec&cz&
$\sigma_{cz}$ & cdE \\
\hline
FCCB~1100  &03:36:41.6  $-$34:59:27  &  19240   &    36    & &  FCCB~1611  &03:42:43.4  $-$36:16:08  &  23108   &    99  & \\ 
FCCB~1103  &03:36:43.0  $-$34:45:34  &  38389   &    65    &*&  FCCB~1618  &03:42:50.0  $-$35:07:57  &  18469   &    74  & \\ 
FCCB~1106  &03:36:47.6  $-$34:32:29  &  19244   &    37    & &  FCCB~1619  &03:42:53.6  $-$34:11:60  &  16161   &    38  & \\ 
FCCB~1107  &03:36:49.7  $-$33:22:46  &  28146   &    51    &*&  FCCB~1621  &03:42:54.1  $-$34:47:51  &  19245   &    71  & \\ 
FCCB~1109  &03:36:53.2  $-$32:59:20  &  35175   &   114    &*&  FCCB~1632  &03:42:54.5  $-$37:28:30  &   6411   &    45  & \\ 
FCCB~1110  &03:36:53.2  $-$33:33:16  &  22083   &    25    & &  FCCB~1626  &03:42:54.6  $-$36:17:56  &  23392   &    98  &* \\ 
FCCB~1116  &03:36:54.0  $-$35:35:52  &  20941   &    76    & &  FCCB~1631  &03:42:54.6  $-$37:10:57  &  15099   &    23  & \\ 
FCCB~1134  &03:37:08.6  $-$36:50:10  &  18833   &    28    & &  FCCB~1635  &03:43:07.1  $-$32:41:35  &  38715   &    77  & \\ 
FCCB~1143  &03:37:17.3  $-$35:09:26  &  18909   &    39    & &  FCCB~1643  &03:43:07.9  $-$34:23:52  &  38289   &    48  & \\ 
FCCB~1144  &03:37:19.9  $-$33:15:20  &  15765   &    43    & &  FCCB~1648  &03:43:08.3  $-$37:07:36  &  26513   &    57  & \\ 
FCCB~1147  &03:37:22.5  $-$33:02:29  &  11702   &    28    &*&  FCCB~1654  &03:43:16.3  $-$33:37:54  &  18228   &    70  & \\ 
FCCB~1156  &03:37:23.2  $-$37:35:08  &  24491   &    34    & &  FCCB~1660  &03:43:21.0  $-$36:18:18  &  12150   &    57  & \\ 
FCCB~1155  &03:37:28.6  $-$33:02:46  &  11884   &    36    & &  FCCB~1663  &03:43:23.1  $-$35:12:24  &  23378   &   102  & \\ 
FCCB~1158  &03:37:32.1  $-$32:26:02  &  31599   &    74    & &  FCCB~1668  &03:43:29.0  $-$33:41:04  &  27643   &    56  & \\ 
FCCB~1173  &03:37:32.8  $-$37:53:28  &  33243   &    59    & &  FCCB~1666  &03:43:29.6  $-$33:27:01  &  22718   &   112  & \\ 
FCCB~1162  &03:37:33.4  $-$32:21:18  &  11813   &    50    &*&  FCCB~1676  &03:43:31.6  $-$36:39:06  &  14054   &    61  & \\ 
FCCB~1171  &03:37:33.5  $-$36:18:26  &  25261   &    69    & &  FCCB~1680  &03:43:33.5  $-$36:18:28  &  25428   &    78  & \\ 
FCCB~1169  &03:37:35.3  $-$34:39:51  &  37055   &    56    & &  FCCB~1684  &03:43:34.1  $-$37:17:03  &  34346   &    68  &* \\ 
FCCB~1178  &03:37:42.7  $-$35:19:11  &  36840   &    42    & &  FCCB~1681  &03:43:35.0  $-$35:48:02  &  30490   &    97  & \\ 
   Name 3  &03:37:45.0  $-$34:48:34  &  26232   &   133    & &  FCCB~1686  &03:43:39.8  $-$34:40:06  &  12759   &    28  & \\ 
FCCB~1186  &03:37:50.1  $-$33:11:57  &  11755   &    68    & &  FCCB~1682  &03:43:40.2  $-$32:37:25  &  19578   &    18  & \\ 
FCCB~1193  &03:37:50.9  $-$34:29:17  &  36748   &   115    &*&  FCCB~1700  &03:43:46.1  $-$37:37:43  &  13214   &    45  & \\ 
FCCB~1195  &03:37:51.4  $-$35:47:35  &  33550   &    63    & &  FCCB~1695  &03:43:47.4  $-$33:54:49  &  52379   &    95  &* \\ 
FCCB~1192  &03:37:52.0  $-$33:36:45  &  15747   &    72    & &  FCCB~1707  &03:43:57.9  $-$36:12:20  &  25083   &    59  & \\ 
FCCB~1200  &03:37:54.9  $-$36:43:52  &  34831   &   157    &*&  FCCB~1714  &03:44:00.6  $-$35:35:28  &  18250   &    69  & \\ 
   Name 4  &03:37:57.1  $-$37:25:43  &  16923   &    31    & &  FCCB~1711  &03:44:04.5  $-$32:41:14  &  49995   &    75  & \\ 
FCCB~1202  &03:38:00.6  $-$33:34:17  &  15879   &    51    & &  FCCB~1727  &03:44:12.0  $-$37:07:34  &  18379   &    87  & \\ 
FCCB~1212  &03:38:02.1  $-$35:59:59  &  36380   &    42    & &  FCCB~1739  &03:44:21.0  $-$35:31:56  &  34769   &    69  & \\ 
FCCB~1209  &03:38:05.9  $-$32:25:32  &  35737   &   123    & &  FCCB~1749  &03:44:29.0  $-$36:56:09  &  13145   &    38  & \\ 
FCCB~1225  &03:38:08.1  $-$36:19:26  &  15477   &    89    & &  FCCB~1754  &03:44:30.1  $-$38:16:49  &  25287   &    55  & \\ 
FCCB~1221  &03:38:10.3  $-$33:17:19  &  32624   &    58    & &  FCCB~1762  &03:44:41.6  $-$35:41:31  &  33718   &    48  & \\ 
FCCB~1230  &03:38:14.0  $-$32:51:40  &  32302   &    64    & &  FCCB~1769  &03:44:44.9  $-$36:38:41  &  22825   &    37  & \\ 
FCCB~1234  &03:38:14.3  $-$33:38:43  &  15929   &    19    &*&  FCCB~1767  &03:44:45.7  $-$34:55:26  &  35093   &    88  & \\ 
FCCB~1238  &03:38:15.9  $-$34:06:55  &  14480   &    40    & &  FCCB~1784  &03:44:51.5  $-$38:34:32  &   6578   &    38  & \\ 
FCCB~1244  &03:38:16.6  $-$37:33:33  &  13743   &    38    & &  FCCB~1797  &03:45:00.6  $-$37:36:25  &  15278   &    37  & \\ 
FCCB~1240  &03:38:18.1  $-$33:07:15  &  32595   &    64    & &  FCCB~1799  &03:45:01.8  $-$38:22:57  &  14684   &    38  & \\ 
FCCB~1257  &03:38:28.3  $-$34:37:49  &  16343   &    53    & &  FCCB~1798  &03:45:04.9  $-$35:30:18  &  35294   &    81  & \\ 
FCCB~1263  &03:38:31.3  $-$35:14:18  &  40600   &    52    & &  FCCB~1810  &03:45:17.4  $-$32:12:48  &  12111   &    73  & \\ 
FCCB~1273  &03:38:36.4  $-$36:04:14  &  18655   &    48    & &  FCCB~1814  &03:45:23.2  $-$35:45:57  &  30418   &    63  &* \\ 
FCCB~1267  &03:38:38.4  $-$31:53:48  &  13263   &    80    & &  FCCB~1816  &03:45:26.6  $-$34:15:45  &  37560   &    38  &* \\ 
FCCB~1271  &03:38:39.2  $-$32:54:37  &  32491   &    60    &*&  FCCB~1848  &03:45:50.1  $-$38:08:35  &  13437   &    51  & \\ 
FCCB~1278  &03:38:42.8  $-$34:20:37  &  16016   &    33    & &  FCCB~1854  &03:45:59.5  $-$34:34:06  &  18268   &    75  & \\ 
FCCB~1276  &03:38:44.3  $-$32:39:18  &  39020   &    57    &*&  FCCB~1857  &03:45:59.9  $-$35:32:14  &  33119   &    55  & \\ 
FCCB~1279  &03:38:45.1  $-$32:47:54  &  32170   &    77    & &  FCC~311    &03:46:18.2  $-$33:45:48  &  47658   &    60  & \\ 
FCCB~1284  &03:38:48.2  $-$32:42:09  &  39398   &    63    & &  FCCB~1892  &03:46:29.1  $-$34:24:21  &  16285   &    31  & \\ 
FCCB~1299  &03:38:52.9  $-$37:30:09  &  18980   &    23    & &  FCCB~1898  &03:46:30.6  $-$34:22:46  &  16130   &    73  & \\ 
FCCB~1303  &03:38:59.1  $-$36:15:03  &  34702   &    79    & &  FCCB~1905  &03:46:32.1  $-$36:53:41  &  18101   &   105  & \\ 
FCCB~1306  &03:39:03.7  $-$35:44:20  &  14838   &    47    & &  FCCB~1903  &03:46:34.6  $-$34:23:01  &  15664   &    61  & \\ 
FCCB~1316  &03:39:08.2  $-$35:52:17  &  25865   &    44    & &  FCCB~1921  &03:46:41.0  $-$36:56:43  &  18258   &    47  &* \\ 
FCCB~1308  &03:39:08.4  $-$33:30:16  &  15698   &    58    &*&  FCCB~1919  &03:46:41.5  $-$35:51:35  &  23558   &    71  & \\ 
FCCB~1315  &03:39:11.6  $-$33:03:15  &  32347   &    97    & &  FCCB~1911  &03:46:43.8  $-$32:48:44  &  35132   &    68  & \\ 
FCCB~1319  &03:39:12.8  $-$34:07:09  &  16002   &    50    & &  FCCB~1932  &03:46:49.3  $-$36:26:45  &  21247   &   116  & \\ 
FCCB~1321  &03:39:14.8  $-$35:15:48  &  18459   &    64    & &  FCCB~1939  &03:46:56.0  $-$35:14:15  &  37814   &   153  & \\ 
FCCB~1320  &03:39:17.7  $-$32:51:27  &  28276   &    57    & &  FCCB~1941  &03:46:56.5  $-$35:46:17  &  25791   &    88  &* \\ 
FCCB~1323  &03:39:20.0  $-$34:27:38  &  21624   &   111    & &  FCCB~1940  &03:46:56.8  $-$35:15:03  &  37596   &    74  & \\ 
FCCB~1335  &03:39:22.8  $-$38:12:12  &  30910   &    53    & &  FCCB~1942  &03:46:58.2  $-$34:47:59  &  27823   &    88  & \\ 
FCCB~1330  &03:39:26.0  $-$34:19:18  &  37032   &    64    &*&  FCCB~1946  &03:47:01.4  $-$34:26:27  &  15993   &   112  &* \\ 
FCCB~1332  &03:39:26.4  $-$34:51:52  &  26789   &    71    & &  FCCB~1951  &03:47:09.2  $-$32:49:22  &  27361   &    65  & \\ 
FCCB~1339  &03:39:31.6  $-$33:06:43  &  32728   &    83    & &  FCCB~1974  &03:47:18.5  $-$38:22:47  &  16325   &   134  & \\ 
FCCB~1348  &03:39:31.6  $-$36:53:18  &  14546   &   106    & &  FCCB~1986  &03:47:35.8  $-$37:29:31  &  23404   &    82  &* \\ 
FCCB~1349  &03:39:38.2  $-$32:37:08  &  38362   &    47    & &  FCCB~1983  &03:47:37.6  $-$34:16:55  &  19188   &    38  & \\ 
FCCB~1353  &03:39:39.0  $-$35:25:14  &  13850   &    37    & &  FCCB~1984  &03:47:38.0  $-$34:21:26  &  16381   &   134  & \\ 
FCCB~1357  &03:39:41.6  $-$33:52:37  &  16356   &    34    & &  FCCB~1987  &03:47:40.0  $-$35:25:03  &  35385   &    55  & \\ 
FCCB~1355  &03:39:43.4  $-$32:25:50  &  29330   &    58    & &  FCCB~1988  &03:47:42.6  $-$35:06:24  &  15255   &    30  & \\ 
FCCB~1378  &03:39:45.9  $-$37:50:41  &  25015   &    66    & &  FCCB~1993  &03:47:44.2  $-$36:08:47  &  23677   &    64  & \\ 
FCCB~1373  &03:39:46.0  $-$36:10:47  &  14693   &    64    & &  FCCB~2007  &03:47:50.8  $-$37:38:31  &  14153   &    69  & \\ 
FCCB~1383  &03:39:52.3  $-$36:59:43  &  42520   &    69    & &  FCCB~2005  &03:47:52.4  $-$36:06:60  &  23180   &    36  & \\ 
FCCB~1377  &03:39:52.4  $-$33:13:14  &  27581   &    33    & &  FCCB~2020  &03:48:06.1  $-$33:50:02  &  33124   &    77  & \\ 
FCCB~1380  &03:39:55.9  $-$32:24:49  &  63367   &   142    & &  FCCB~2032  &03:48:15.5  $-$34:30:14  &  15806   &    48  & \\ 
FCCB~1384  &03:39:56.5  $-$35:53:16  &  35941   &    42    & &  FCCB~2037  &03:48:16.4  $-$36:05:25  &  30270   &   118  & \\ 
FCCB~1396  &03:40:02.1  $-$37:47:31  &  13809   &    55    & &  FCCB~2050  &03:48:28.9  $-$33:19:44  &  35242   &    57  &* \\ 
FCCB~1389  &03:40:03.8  $-$32:56:53  &  17794   &    55    & &  FCCB~2047  &03:48:29.2  $-$32:05:20  &  21665   &    62  & \\ 
FCCB~1404  &03:40:05.2  $-$37:47:32  &  13800   &    49    & &  FCCB~2051  &03:48:30.8  $-$32:14:21  &  21731   &    62  & \\ 
FCCB~1399  &03:40:05.6  $-$36:17:12  &  40193   &    45    & &  FCCB~2058  &03:48:34.9  $-$33:43:55  &  16005   &    12  & \\ 
FCCB~1406  &03:40:07.1  $-$37:08:59  &  24391   &    45    &*&  FCCB~2064  &03:48:39.3  $-$34:02:48  &  49439   &    47  & \\ 
FCCB~1407  &03:40:09.1  $-$36:05:58  &  25652   &    70    & &  FCCB~2063  &03:48:39.5  $-$33:44:13  &  16193   &    27  & \\ 
FCCB~1411  &03:40:10.4  $-$37:49:40  &  14099   &    47    &*&  FCCB~2067  &03:48:40.9  $-$35:54:45  &  18764   &    33  & \\ 
FCCB~1408  &03:40:13.6  $-$33:43:42  &  22104   &    48    & &  FCCB~2087  &03:48:56.5  $-$35:50:59  &  23125   &   104  & \\ 
FCCB~1421  &03:40:19.5  $-$37:14:17  &  14241   &    82    & &  FCCB~2080  &03:48:57.2  $-$32:37:17  &  22075   &    53  & \\ 
FCCB~1419  &03:40:24.7  $-$32:34:49  &  28579   &    47    & &  FCCB~2092  &03:49:02.2  $-$35:20:57  &  12181   &    31  & \\ 
FCCB~1430  &03:40:27.4  $-$37:49:56  &  12983   &    44    & &  FCCB~2098  &03:49:02.6  $-$36:26:08  &  34292   &   152  & \\ 
FCCB~1436  &03:40:30.4  $-$37:49:40  &  12855   &    39    & &  FCCB~2096  &03:49:03.5  $-$35:36:21  &  18103   &    46  & \\ 
FCCB~1426  &03:40:32.7  $-$33:05:13  &  21844   &    59    &*&  FCCB~2111  &03:49:11.9  $-$36:33:40  &   5679   &    35  & \\ 
FCCB~1446  &03:40:42.8  $-$33:55:47  &  15986   &    39    & &  FCCB~2138  &03:49:43.7  $-$36:22:20  &  34835   &    63  & \\ 
FCCB~1452  &03:40:49.7  $-$35:33:24  &  14669   &    87    & &  FCCB~2174  &03:50:22.2  $-$33:56:53  &   7096   &    47  & \\ 
FCCB~1454  &03:40:53.1  $-$34:25:28  &  52838   &    41    & &  FCCB~2180  &03:50:29.4  $-$34:15:13  &  38068   &   135  & \\ 
FCCB~1456  &03:40:55.8  $-$33:10:01  &  42490   &    97    & &  FCCB~2203  &03:50:44.4  $-$37:22:32  &  23150   &    86  & \\ 
FCCB~1467  &03:41:03.8  $-$34:15:15  &  36937   &    74    & &  FCCB~2197  &03:50:45.1  $-$35:56:27  &  23107   &    32  & \\ 
   Name 5  &03:41:11.5  $-$36:45:50  &  31302   &    36    & &  FCCB~2199  &03:50:49.2  $-$33:33:59  &  27597   &   127  & \\ 
FCCB~1487  &03:41:13.9  $-$37:45:05  &  34196   &    59    & &  FCCB~2225  &03:51:02.0  $-$36:01:39  &  18100   &    22  & \\ 
FCCB~1498  &03:41:21.1  $-$36:41:01  &  15170   &    28    & &  FCCB~2230  &03:51:11.9  $-$33:50:10  &  19380   &    77  & \\ 
FCCB~1523  &03:41:35.0  $-$36:16:50  &   6794   &   102    & &  FCCB~2234  &03:51:14.1  $-$34:25:27  &  16313   &    43  &* \\ 
FCCB~1520  &03:41:36.5  $-$33:40:58  &  21744   &    50    & &  FCCB~2235  &03:51:15.3  $-$33:49:02  &  19285   &    43  & \\ 
FCCB~1528  &03:41:36.5  $-$38:00:53  &  13664   &    45    & &     Name 6  &03:51:17.1  $-$35:45:49  &  18102   &    51  & \\ 
FCCB~1538  &03:41:42.2  $-$37:55:29  &  13681   &    38    & &  FCCB~2239  &03:51:21.0  $-$32:46:00  &  22340   &    47  &* \\ 
FCCB~1532  &03:41:43.0  $-$34:07:26  &  12161   &    62    & &  FCCB~2248  &03:51:22.3  $-$37:07:40  &  22473   &    42  & \\ 
FCCB~1533  &03:41:43.3  $-$34:29:41  &  37316   &    43    & &  FCCB~2251  &03:51:24.9  $-$37:20:22  &  16669   &    32  &* \\ 
FCCB~1541  &03:41:48.2  $-$36:48:51  &   8377   &    69    &*&  FCCB~2263  &03:51:37.4  $-$34:54:11  &  18466   &    35  & \\ 
FCCB~1548  &03:41:57.2  $-$32:45:08  &  11534   &    64    & &  FCCB~2276  &03:51:50.1  $-$37:36:25  &  19162   &    39  & \\ 
FCCB~1562  &03:41:59.9  $-$38:01:57  &  13540   &    38    & &     Name 7  &03:51:59.5  $-$36:14:20  &  14505   &    34  & \\ 
FCCB~1577  &03:42:06.6  $-$36:38:20  &  15426   &    57    & &  FCCB~2289  &03:52:13.2  $-$33:25:09  &  22390   &   181  &* \\ 
FCCB~1594  &03:42:32.6  $-$33:12:17  &  19474   &    57    & &  FCCB~2290  &03:52:16.4  $-$32:07:22  &  10083   &    62  &* \\ 

\hline
\end{tabular}}

Notes: the names of the galaxies not listed in the FCC are given in
Table~\ref{tab_not}. The asterisks indicate cdE candidates from the FCC.
\end{table*}
%%%%%%%%%%%%%%%%%%%%%%%%%%%%%%%%%%%%%%%%%%%%%%%%%%%%%%%%%%%%%%%%%%%%%%%

\end{document}